\documentclass[runningheads]{llncs}
\usepackage{caption}
\usepackage{subcaption}
\usepackage[T1]{fontenc}
\usepackage{graphicx}
\usepackage{comment}
\usepackage{bbding}

\usepackage{url}            
\usepackage{multirow}
\newcommand{\nop}[1]{}

\usepackage{amssymb}
\usepackage{graphicx}
\usepackage{booktabs}
\usepackage{multirow}
\usepackage{lipsum}
\usepackage{pifont}
\usepackage{caption}
\usepackage[utf8]{inputenc}

\usepackage{makecell}

\begin{document}
\title{Reducing Communication Overhead in the IoT-Edge-Cloud Continuum: A Survey on Protocols and Data Reduction Strategies}
\titlerunning{Reducing Communication Overhead in the IoT-Edge-Cloud Continuum}
%
\author{Dora Kreković\inst{1}\and
Petar Krivić\inst{1} \and
Ivana Podnar Žarko\inst{1} \and
Mario Kušek\inst{1} \and
Danh Le-Phuoc\inst{2}}
\authorrunning{Kreković, Krivić, Podnar Žarko, Kušek, Le-Phuoc}
%
\institute{University of Zagreb Faculty of Electrical Engineering and Computing \\
	 Unska 3, 10000 Zagreb, Croatia. \\ \email{dora.krekovic@fer.hr, petar.krivic@fer.hr, ivana.podnar@fer.hr, mario.kusek@fer.hr}\\ 
\and
Technische Universität Berlin \\Straße des 17. Juni 135, 10623 Berlin, Germany
\email{danh.lephuoc@tu-berlin.de}
} 
\maketitle              
\begin{abstract}
The adoption of the Internet of Things (IoT) deployments has led to a sharp increase in network traffic as a vast number of IoT devices communicate with each other and IoT services through the IoT-edge-cloud continuum. This network traffic increase poses a major challenge to the global communications infrastructure since it hinders communication performance and also puts significant strain on the energy consumption of IoT devices. To address these issues, efficient and collaborative IoT solutions which enable information exchange while reducing the transmitted data and associated network traffic are crucial. This survey provides a comprehensive overview of the communication technologies and protocols as well as data reduction strategies that contribute to this goal. First, we present a comparative analysis of prevalent communication technologies in the IoT domain, highlighting their unique characteristics and exploring the potential for protocol composition and joint usage to enhance overall communication efficiency within the IoT-edge-cloud continuum. Next, we investigate various data traffic reduction techniques tailored to the IoT-edge-cloud context and evaluate their applicability and effectiveness on resource-constrained and devices. Finally, we investigate the emerging concepts that have the potential to further reduce the communication overhead in the IoT-edge-cloud continuum, including cross-layer optimization strategies and Edge AI techniques for IoT data reduction.
The paper offers a comprehensive roadmap for developing efficient and scalable solutions across the layers of the IoT-edge-cloud continuum that are beneficial for real-time processing to alleviate network congestion in complex IoT environments.

\end{abstract}
\section{Introduction}
\label{sec:introduction}
\begin{figure}
    \centering
    \includegraphics[width=0.85\linewidth]{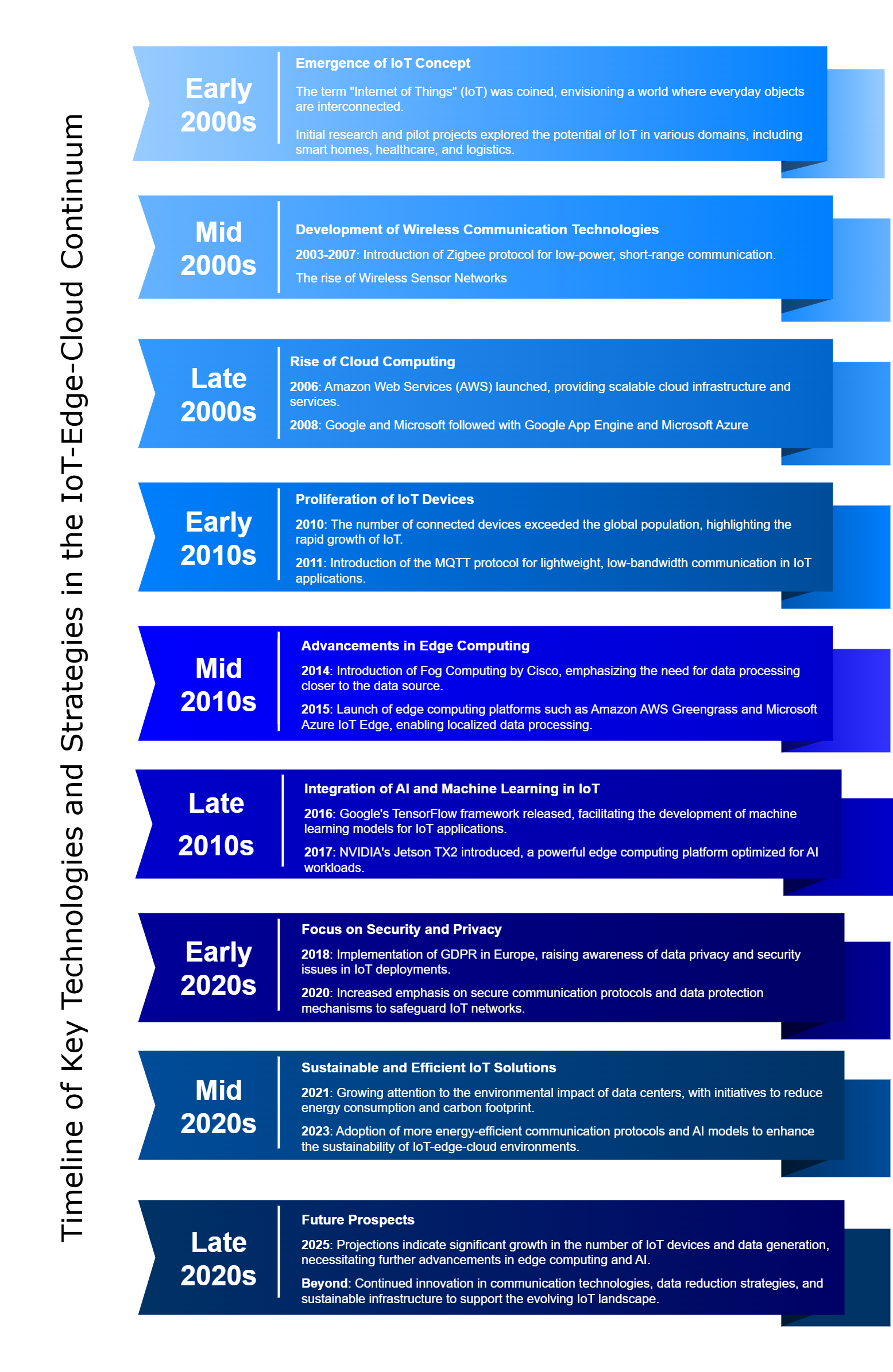}
    \caption{Evolution of key technologies and strategies relevant to the IoT-edge-cloud continuum}
    \label{fig:evolution}
\end{figure}

Modern Internet of Things (IoT) solutions are  becoming increasingly distributed and even decentralized, operating within heterogeneous and complex computing and communication environments. Large-scale IoT deployments are composed of numerous sensors and actuators spread across diverse locations which generate vast amounts of data, creating a need for alternative architectures that move processing to the edge. 
This move away from centralized cloud-based IoT platforms has led to the concept of the \textit{IoT-edge-cloud continuum}. With traditional, centralized approaches, the majority of IoT traffic is sent to cloud servers for processing and storage, resulting in network congestion and long response times. According to a report by Cisco \cite{cisco}, data transmission costs can constitute a significant portion of operational expenses in IoT deployments, especially in scenarios involving large data volumes. Additionally, the energy consumption required for transmitting, processing, and storing data in cloud servers represents a substantial component of overall operational costs within IoT environments. 
It has been shown \cite{energyiot} that up to 80\% of the total energy consumption in IoT sensor networks is due to wireless data transmission. As the number of IoT devices proliferates and data generation continues to grow exponentially, projections indicate a significant future increase in energy consumption by data centers worldwide \cite{iea}.
The IoT-edge-cloud continuum reverses this trend by enabling data processing as close to data source as possible – on nearby edge devices or even directly on the IoT devices themselves. This significantly reduces the overall network traffic to the cloud and leads to a more responsive and efficient IoT system for processing data and reacting to events in local smart environments hosting various IoT devices. The timeline and evolution of key technologies and strategies relevant to the IoT-edge-cloud continuum is presented in Figure \ref{fig:evolution}.

The IoT-edge-cloud continuum is facilitated by \textit{edge computing}: It introduces an additional layer within the IoT architecture, positioned between IoT devices and the cloud \cite{7469991}. In this setup devices can make real-time decisions with reduced latency, which is critical for applications requiring immediate action, such as industrial automation and autonomous vehicles. Additionally, pre-processing and filtering of raw IoT data at the edge reduces the amount of data transmitted within the continuum and the overall data volume sent to the cloud, thereby conserving network bandwidth.  This is particularly beneficial in IoT scenarios with limited bandwidth or high data volumes. Finally, edge computing enhances security by allowing sensitive data to be processed and analyzed locally, minimizing the risk of data exposure during transmission to the cloud.

The sheer volume of data generated by IoT devices can overwhelm network resources, especially considering the nature of \textit{big data}. According to \cite{big_data}, big data is characterized by large, uncertain, and volatile volumes, often in unstructured formats that are challenging for traditional computing systems to process. In this context, data reduction strategies play a crucial role in mitigating this challenge by minimizing the amount of data transmitted or stored while preserving essential information, quantified in terms of bits per unit of transmitted information. These strategies include \textit{compression}, which reduces the size of data through encoding techniques; \textit{prediction}, which anticipates future data based on past patterns to transmit only necessary updates; and \textit{aggregation}, which consolidates multiple data points into a single representation. The optimal combination of communication protocol and data reduction technique depends on the specific requirements of the application and the capabilities of the devices involved.

This survey paper aims to answer the following research questions:

    \begin{enumerate}
        \item Which \textit{communication technologies and protocols} are most suitable for different aspects of the IoT-edge-cloud continuum? This includes analyzing both wireless communication technologies and application-layer protocols, considering factors such as power consumption, range, throughput, and data overhead.
        \item Which \textit{data reduction strategies} (compression, prediction, aggregation) are most effective in reducing data transmission within the continuum? The analysis considers the trade-offs between data accuracy and processing complexity.
        \item How can we effectively map specific communication protocols and data reduction strategies to the different layers of the IoT-edge-cloud continuum, taking into account the limitations of resource-constrained edge devices?
    \end{enumerate}

IoT solutions require efficient data transmission to facilitate real-time communication and collaboration between devices and services. The primary goal of this survey is to analyze the available options to reduce the communication overhead within the IoT-edge-cloud continuum in this context. By analyzing suitable communication protocols and data reduction strategies, we aim to provide a comprehensive guide for researchers and practitioners to design efficient and scalable IoT solutions that are particularly valuable for applications requiring real-time processing and minimal network congestion in complex IoT environments. Scalability, in this context, refers to the capability of IoT systems to integrate an increasing number of devices and data sources without compromising performance or requiring significant redesign. This includes accommodating growth in both the number of connected devices and the volume of data generated, ensuring that the system can maintain its effectiveness as it scales.

\subsection{Comparison with related surveys}
Existing surveys, as summarized in Table~\ref{tab:survey}, have primarily focused on either data reduction techniques or communication technologies in isolation. 
This survey fills this gap by providing a comprehensive analysis of both aspects: We analyze both the data reduction techniques for reducing data size before transmission, and characteristics of widespread IoT protocols and technologies so that the combination can ensure efficient data transmission in the IoT-edge-cloud continuum.

\begin{table}[h]
  \centering
  \footnotesize
  \caption{Comparison of relevant surveys.}
  \resizebox{\textwidth}{!}{
  \begin{tabular}{cccccc}
    \toprule
    \textbf{Work}& \textbf{Year}& \textbf{Data reduction techniques}& \textbf{Communication technologies}& \textbf{IoT Context} & \textbf{Focused on the Edge} \\
    \midrule
    \cite{9063670} & 2020 & x & \ding{51} & \ding{51} &  \ding{51}\\
    \midrule
    \cite{TALAVERA2017283} & 2017 & x & \ding{51} & \ding{51} & x\\
    \midrule
    \cite{inproceedings} & 2017 & \ding{51} & x & x & x\\
    \midrule
    \cite{article} & 2022 & \ding{51} & x & \ding{51} & \ding{51}\\
    \midrule
    \cite{GERODIMOS20231} &2023 & x & \ding{51} & \ding{51} & x\\
    \midrule
    \cite{NASSRA2023100806}  &2023 & \ding{51} & \ding{51} & \ding{51} & x \\
    \midrule
    \cite{aggregation_survey} &2023 & \ding{51} & \ding{51} & \ding{51} & x\\
 
  \end{tabular}
  }
   \label{tab:survey}
\end{table}

Several surveys explore data transmission within specific IoT application domains. For instance, \cite{9063670} investigates how edge computing can accelerate the development of smart cities and provides an overview of relevant literature to propose a taxonomy categorizing edge computing applications in this context. Similarly, \cite{TALAVERA2017283} focuses on the technological solutions for IoT in agro-industrial and environmental sectors, explaining the underlying infrastructure and technologies.

Other surveys analyze data reduction within a broader big data context. In~\cite{inproceedings}, the authors looked at the workflow of data preprocessing in the context of big data. They provided insights into the four phases of data pre-processing, including cleansing, integration, reduction and transformation.

Some surveys, e.g.~\cite{GERODIMOS20231}, emphasize the aspect of communication technologies in the IoT. In this work, the authors deal with the basics of IoT architecture and give an overview of the communication protocols developed specifically for IoT solutions. The authors also analyze security threats and general implementation challenges.

In~\cite{article}, a comprehensive overview of the data reduction techniques applicable at the edge is presented. The review covered literature on data reduction solutions at the edge of IoT systems. Algorithms, hardware technologies, the data types used and the objects that contribute to data reduction were analyzed and discussed.

Some of the studies listed in Table~\ref{tab:survey} focus on a particular method or reduction technique and provide its detailed analysis. For instance, the authors in \cite{NASSRA2023100806} focus exclusively on the use of data compression (DC) as a technique for reducing data traffic specifically in Wireless Body Sensor Networks (WBSN). Further, authors in~\cite{aggregation_survey} presented an in-depth survey of data aggregation (DA) protocols based on their ability to address issues related to topology, interference, fault-tolerance, security and mobility.

While all the referenced surveys exhibit depth and comprehensiveness, our main contribution lies in examining jointly data reduction strategies and communication protocols within the scope of the IoT-edge-cloud continuum.

\subsection{Our contribution}
Traditional approaches to IoT communication can lead to 1) significant degradation of communication performance due to high data volumes which can overwhelm network resources, causing congestion and delays in data transmission and 2) increased energy consumption of IoT devices due to frequent data transmissions.
This survey addresses these challenges by focusing on efficient and collaborative communication in the IoT-edge-cloud continuum by investigating jointly the available methods for reducing the data generated by IoT devices and communication protocols contributing to reduced network traffic.

Our contribution can be summarized as follows:
 \begin{itemize}
     \item We provide a detailed comparison of the prevailing communication technologies in the IoT domain.  This analysis goes beyond simply identifying their characteristics.  We investigate how the combined use of protocols at different layers of the IoT-edge-cloud continuum can enhance the overall communication efficiency.
     \item We explore various techniques for data reduction and in combination with communication technologies to critically evaluate their applicability within the context of the IoT-edge-cloud continuum.
     \item We investigate the emerging concepts that have the potential to further reduce the communication overhead in the IoT-edge-cloud continuum, including cross-layer optimization strategies and Edge AI techniques.
 \end{itemize}
Through a combined analysis of communication protocols, data reduction techniques, and emerging concepts, this survey provides a comprehensive understanding of methods for reducing communication overhead in the evolving IoT-edge-could continuum, leading to the development of scalable and energy-efficient IoT systems in complex environments.

The paper is structured as follows: Section~\ref{sec: transTech} provides an overview of prevalent wireless communication technologies and application layer protocols commonly used in the IoT-edge-cloud continuum. Additionally, we compare the efficiency of these communication protocols to identify their suitability for modern IoT use cases. In Section \ref{sec: dataRed}, we explore various data reduction strategies, discussing their advantages and disadvantages, identify their placement within the continuum, and discuss their usage in combination with appropriate communication technology. We also conduct a comparative analysis of data reduction strategies. Section \ref{sec:challenges} addresses the emerging concepts and open challenges concerning data traffic reduction strategies in complex IoT environments. Finally, Section \ref{sec:conclusion} concludes the paper, summarizing key research observations and outlining future research directions.

\section{Communication technologies for the IoT-edge-cloud continuum} \label{sec: transTech}

\begin{figure}[ht]
\centerline{\includegraphics[width=0.95\linewidth]{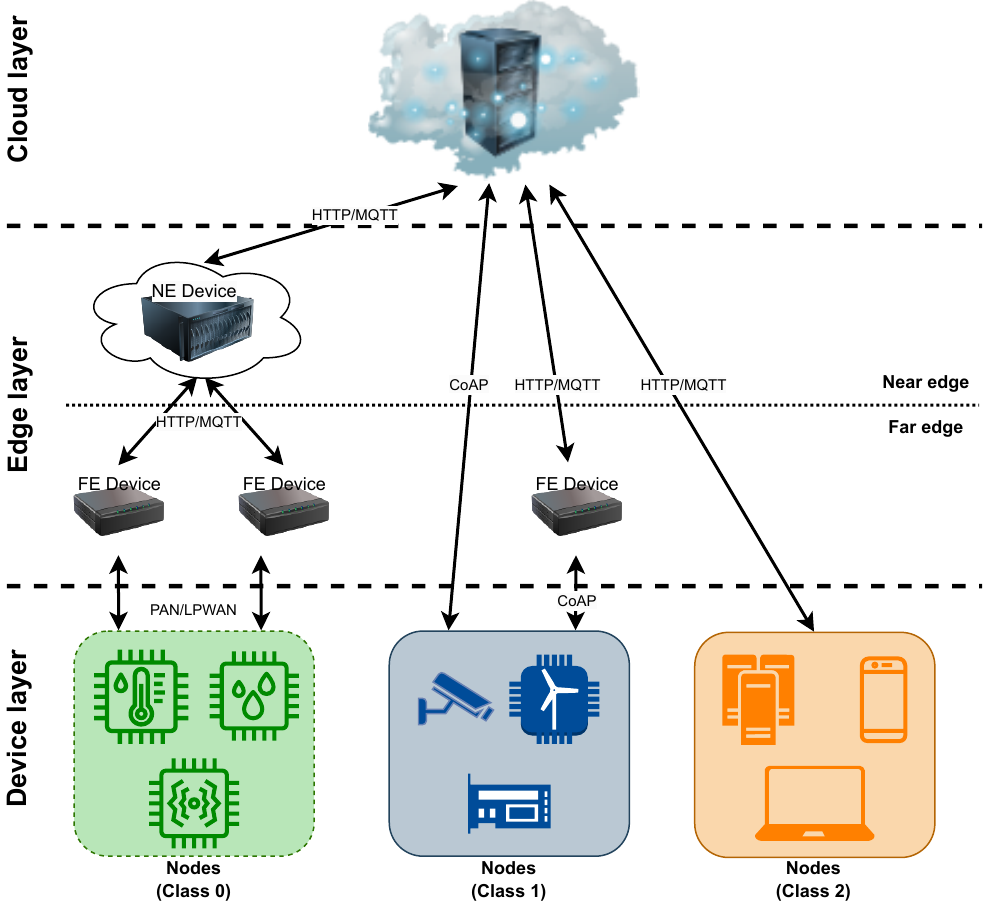}}
\caption{Layered architecture of the IoT stack with edge computing layer.}
\label{fig:edgeStack}
\end{figure}

An IoT environment typically utilizes a hierarchical three-layered general architecture depicted in Figure \ref{fig:edgeStack} consisting of the device layer, edge layer and cloud layer forming the IoT-edge-cloud continuum. Data generated at the device layer is forwarded upwards to the cloud, either directly or through edge devices. Conversely, actuation commands and processed data flow from the cloud back to devices. This data transmission across layers relies on various wireless communication technologies and application-layer protocols, chosen based on the characteristics of the devices and services running within the environment. The heterogeneity of IoT environments is evident in diverse IoT devices employed at the bottom layer, ranging from wearables and agricultural sensors to robots and autonomous vehicles as well as computing nodes with different capabilities positioned at the edge layer, ranging from, e.g.,  Raspberry Pi (RPi) and Jetson Nano/Orin to a local micro-cloud with a few server racks. 

\begin{table}[ht]
\caption{IoT Device Categories}
\label{tab:iot_categories}
\begin{tabular}{|l|p{3cm}|l|p{8cm}|} 
\hline
\textbf{Category} & \textbf{Resources} & \textbf{Protocol} & \textbf{Description} \\ \hline
Class 0 & \textless10 KB RAM, \textless100 KB Flash, battery-powered & LPWA & Devices with severe constraints, such as 8-bit or 16-bit microcontrollers, typically utilize LPWA communication protocols for transmitting data to gateways. Gateways function as edge nodes, converting data into compatible formats before relaying it to the cloud. Due to their limitations, these devices are unable to run the complete IP stack.\\ \hline
Class 1 & $\sim$10 KB RAM, $\sim$100 KB Flash & CoAP & Devices with moderate resources, unable to employ a full IP stack. They utilize constrained application protocols like CoAP to interact with the cloud and edge nodes directly without requiring gateways. \\ \hline
Class 2 & \textgreater50 KB RAM, \textgreater250 KB Flash & Full IP stack & Devices with ample resources capable of running full implementations of the IP stack. They can directly connect to the cloud and edge nodes, but consume more energy compared to Class 0 and Class 1 devices. \\ \hline
\end{tabular}
\end{table}

IoT devices fall into three categories of constrained devices (Class 0, 1 and 2) specified in RFC 7228 \cite{rfc7228} as shown in Table \ref{tab:iot_categories}. Nodes in Class 0 are severely constrained (8-bit or 16-bit microcontrollers, $<< 10 KB$ of RAM, $<<100 KB$ of Flash, battery powered) and typically use a constrained communication protocol (Low-Power Wide-Area, LPWA) for data transmission to gateways, since they do not have sufficient resources to run the complete IP stack. The gateways act as edge nodes and translate the received data into a format compatible with the chosen application-layer protocol before forwarding it to the cloud. The selection of this application-layer protocol depends on factors such as requirements of a specific use case, resources available at the gateway, and the cloud service receiving the data. 

For devices communicating directly with an IoT platform in the cloud, RFC 7228 categorizes them as Class 1 or Class 2 nodes \cite{rfc7228}). Class 2 nodes have more resources ($> 50KB$ of RAM, $>250 KB$ of Flash) and can run full implementations of the IP stack,  enabling direct connection to the cloud and edge nodes at the expense of higher energy consumption. Class 1 devices offer a middle ground. Possessing more resources than Class 0 (approximately 10KB RAM and 100KB Flash), they cannot employ a full IP stack but leverage constrained application protocols like CoAP to interact with the cloud and edge nodes without requiring a gateway.   

The IoT-edge-cloud continuum incorporates additional classes of devices within the edge layer, which is further divided into the far edge and near edge environment (Figure \ref{fig:edgeStack}). Notably, edge nodes within the category of far edge (FE) nodes can also be classified as Class 2 devices, while edge nodes within the category of near edge (NE) nodes possess even greater processing power than FE nodes. These devices, both FE and NE nodes, often operate under the following constraints:
    \begin{itemize}
        \item \textbf{Limited processing power}: Edge devices typically have less processing capability than cloud servers, restricting the complexity of tasks they can handle at the edge.
        \item \textbf{Constrained memory}: Data storage capacity on edge devices is often limited, requiring careful consideration of data processing and storage strategies.
        \item \textbf{Limited energy source}: Battery life is critical for many FE devices and communication protocols that minimize energy consumption should be employed.
        \item \textbf{Network diversity}: Edge devices can utilize various communication modules. Some support LPWA protocols which are ideal for transmitting small data packets over long distances in IoT environments. Others connect via WiFi, which offers higher bandwidth but shorter range and higher power consumption.
    \end{itemize}

In this complex environment, diverse use cases are deployed: for example, continuous environmental monitoring \cite{liu2024multiobjectiveoptimizationdatacollection} or human well-being tracking, or efficient interaction between autonomous vehicles and the surrounding infrastructure\cite{9776521}. Due to the different requirements of such diverse use cases, various communication technologies and protocols are defined for efficient data transmission. Furthermore, the selected protocols need to be in line with the specific hardware employed within each layer of the three-tier architecture. When selecting a communication technology for data transmission between constrained devices and their first-hop neighbors, factors such as \textit{communication range}, \textit{frequency band}, \textit{delay}, \textit{energy consumption}, \textit{battery life}, and \textit{sleep mode configuration} must be considered. The choice of a suitable protocol at the application layer depends on the communication architecture within a specific use case to enable efficient data transfer from the device or edge layer to the cloud layer.   

The application-layer protocols commonly used in IoT environments often introduce overhead by adding fixed-size headers to each transmitted packet. This can be particularly detrimental in scenarios where small amounts of data are frequently transmitted (e.g. a few sensor readings), a common occurrence in IoT communication. Therefore, a thorough analysis of the specific use case is crucial for selecting an optimal application-layer protocol that efficiently utilizes available data buffers within each packet and minimizes overhead while effectively transmitting data.

\subsection{Currently prevalent wireless communication technologies for IoT}
In existing IoT setups, different wireless communication technologies are used to connect end devices with gateway nodes connected to the Internet. These technologies were mostly determined based on range, bandwidth, coverage, power consumption, reliability, latency and cost-effectiveness required within a specific use case \cite{9066337}. In the design phase of each IoT solution, it is necessary to conduct a detailed analysis of the specific IoT use case and properly prioritize the aforementioned features to choose an optimal communication protocol for the particular scenario. In continuation, we briefly introduce technologies that are most commonly used in existing IoT scenarios:

\begin{description}
   \item [BLE] (Bluetooth Low Energy) is a Wireless Personal Area Network (WPAN) technology
   originally designed for short-range communication with smart IoT sensors and devices with limited power consumption \cite{9706334}. Today, it is the most widespread WPAN protocol in IoT, as it is supported by computing devices such as mobile phones and laptops as well as various peripherals. BLE is a special version of standard Bluetooth that has been available since the Bluetooth 4.0 specification and is designed for short-range data exchange with minimal power consumption. Since its emergence, BLE has become the preferred choice over standard Bluetooth in the context of IoT, as it is specifically optimized for energy-constrained communication, a crucial factor for devices in this layer of the IoT stack. The main advantage of BLE compared to its counterparts is that it is more energy-conservative than WiFi and ZigBee, which are usually considered together due to their similarities.
   \item[ZigBee] is a WPAN protocol mainly used in smart home applications \cite{ZOHOURIAN2023100791}. Its main advantage is lower power consumption, as it was originally developed to support low-cost and low-power connectivity for devices that require battery life of several months to several years, but do not require data transfer rates as high as  Bluetooth \cite{ergen2004zigbee}. Although BLE has significantly reduced the power consumption of Bluetooth to an even lower level than ZigBee \cite{6616827} with a lower data rate due to the use of the 802.15.4 standard, ZigBee remains a common choice for many IoT use cases. The reason for this is that most IoT applications do not require a high data rate, and ZigBee still offers a greater range than BLE.
   \item[WiFi] is a wireless communication protocol used to link devices in range of each other and to connect them to the Internet through the wireless router. As it was not originally designed for IoT use cases, but to connect conventional devices such as laptops, smartphones, and servers, some of its key aspects should be revised in order to make it suitable for dense IoT environments \cite{8847082}. Its advantage over BLE and ZigBee  is a considerably higher data rate and a direct connection to the Internet, while its main disadvantage is high power consumption.
   \item[LoRaWAN] is a LPWA network (LPWAN) technology that connects IoT devices over a wider area, making it more suitable for use cases such as smart agriculture. It offers the possibility for private network deployments and easy integration with a range of global network platforms. For these reasons, and due to its open access specification, LoRaWAN has attracted the attention of the research community since it first appeared on the market \cite{s18113995}. It employs a star topology architecture, where end devices send messages to a LoRaWAN gateway, which encapsulates the received messages in UDP/IP packets and forwards them to the network server, while communication of commands towards the end devices takes place in the opposite direction. Its downside compared to similar communication technologies is  low data rate and lower reliability. However, since it operates in an unlicensed spectrum, LoRaWAN offers the possibility to set up a simple private wide area network with a range of up to 20 km, which makes it a popular choice for research projects and startup phases of IoT use cases, where transmission reliability is not critical, and the scale of a solution is not too big to be managed without third-party telecom companies.
   \item[NB-IoT] is the leading LPWAN communication technology  operating in a licensed spectrum and consequently offers advantages over LoRaWAN in terms of data rate, latency, reliability, and range at the expense of specific benefits that LoRa offers by operating in an unlicensed spectrum - battery lifetime, capacity, and cost \cite{SINHA201714}. It was developed by 3GPP, and provides access to network services using a physical layer optimized for very low power consumption and low cost \cite{7962670}. NB-IoT devices are attached to base stations, but the technology does not support authentication and certificate transfer between different cells: The devices must manually select the base stations \cite{MOHAN2023103723} as they use a power saving mode (PSM) where they can turn off their radio components but remain registered to an LTE network. This consequently prevents handovers and, therefore, also the mobility of the connected devices \cite{10.1145/3551663.3558596}. 
   Nevertheless, NB-IoT is a viable choice for use cases that require the abovementioned benefits, especially in dense urban environments where LoRa performance degrades significantly.
   
   \item[Cellular] communication technologies imply 2G/3G, 4G, and 5G mobile networks that were not primarily intended for IoT use cases. However,
   the 5G cellular network has evolved to become a key IoT enabler for three main groups of use cases: enhanced mobile broadband (eMBB), massive machine-type communications (mMTC), and ultra-reliable low-latency communications (URLLC) \cite{9711564}. To support such use cases, the KPI targets for the 5G network were demanding in terms of peak data rate, spectral efficiency, latency, reliability, energy efficiency, connection density, and mobility. Hence, cellular technologies are more robust and offer better performance on most of the considered attributes compared to the previously mentioned  technologies, with the exception of cost and energy efficiency, which are significant factors for most IoT use cases. However, as eMBB and URLLC use cases in IoT prevail over the currently most popular mMTC group, 5G could become an important technology to support IoT use cases for reliable and efficient communication with minimal latency.  
\end{description}

\begin{table}
\centering
\caption{Comparison of currently prevalent communication technologies in IoT.}
\begin{tabular}{|c|c|c|c|c}
\cline{1-4}
         & Range  & Power & Data rate &  \\ \cline{1-4}
BLE      & < 100 m \cite{BLErange}  &  1   &    < 1-2 Mbps       &  \\ \cline{1-4}
ZigBee   & indoor: 25 m, outdoor: 60 m \cite{ZigBeeRange}  &   4    &  < 250 Kbps   &  \\ \cline{1-4}
WiFi     & indoor: <15 m, outdoor: <450 m \cite{WiFiRange}  &   5    &  < 2.4 Gbps   &  \\ \cline{1-4}
LoRaWAN  & urban: 2-5 km, suburban: 15 km \cite{LoRaWANRange} &   2    &  < 50 Kbps        &  \\ \cline{1-4}
NB-IoT   & within the range of a base station &   3    &   < 250 Kbps        &  \\ \cline{1-4}
Cellular & within the range of a base station &  6     &   < 1 Gbps       &  \\ \cline{1-4} 
\end{tabular}
\label{tableProtocols}
\end{table}

Table \ref{tableProtocols} compares the described protocols with regard to the attributes that solution architects should consider during the design phase of a particular IoT setup: range, power consumption and data rate. Range is an attribute that depends on many factors, such as barriers between communicating devices, interference, module manufacturers, etc. Thus, the values for the same protocol differ across the literature, but the ratio between the compared technologies remains exact \cite{BLErange,ZigBeeRange,WiFiRange,LoRaWANRange}. It confirms that BLE, ZigBee and WiFi are WPAN protocols with a shorter range than LPWAN technologies such as LoRaWAN and NB-IoT.

Power consumption is a parameter that cannot be specified precisely. Thus, we use categories (from 1 as the lowest to 6 as the highest) to characterize protocols in terms of power consumption since the ratio between the compared technologies is consistent in the literature \cite{4460126,8053135,7334098,articlewbfb,9068491}. In Table \ref{tableProtocols}, we rated the communication technologies from one for BLE, which is the most energy-efficient, to six for cellular communication, which consumes the most energy. From the reported values, it is clear that BLE, ZigBee, LoRaWAN and NB-IoT are protocols designed primarily for constrained IoT devices, as they consume significantly less energy than WiFi and cellular communication that are primarily used by devices with a steady power supply.

Data rate is an important parameter for IoT deployments since it determines the speed of data transfer between devices using a specific communication protocol. The values given in Table \ref{tableProtocols} can be found in the literature and imply the maximum expected uplink data rate, which is rarely achieved in practice \cite{8614763,5942102,fi11010016,SINHA201714}. However, the values clearly outline that WiFi and cellular communication offer significantly higher data rates than other  protocols, which is one of the reasons for their higher energy consumption. BLE offers a slightly lower data rate than WiFi and cellular, but still has an advantage over the other protocols developed specifically for IoT - ZigBee, LoRaWAN and NB-IoT. BLE can be used to connect computer peripherals that require higher data rates than the IoT data sent from the sensors, which usually come in small packets that ZigBee, LoRaWAN and NB-IoT can also support.

In the context of data traffic reduction, the main motivation is to reduce the overall data traffic through the global Internet, while strategies for network traffic reduction over WPAN and LPWAN links are not that critical, but should not be neglected since by reducing the number of transmissions on a data-link layer, efficiency of sensor devices increases. Hence, the most energy-efficient wireless communication technology that offers a data rate and range required by a specific use case is the best choice for establishing a connection between IoT devices and the gateway. Further data reduction of the Internet traffic is then based on different strategies that can be applied at the gateway and edge level in line with the used application layer protocols which we analyze in the following subsection.

\subsection{Application layer protocols in IoT}

A comprehensive overview of application layer protocols for IoT environments and their recent advances is presented in \cite{DONTA2022727}. However, not all protocols presented (e.g. XMPP and WebSocket) are equally suitable for an IoT environment. Thus, we only consider the protocols that most commonly used in IoT deployments today. Application-layer protocols can be classified as either request-response or publish-subscribe protocols, depending on the communication model used. We briefly introduce two of the most common options for each group. Protocols implementing the request-response communication model include HTTP (\textit{Hypertext Transfer Protocol}) and CoAP (\textit{Constrained Application Protocol}), while protocols implementing the publish-subscribe model are MQTT (\textit{Message Queue Telemetry Transport}) and AMQP (\textit{Advanced Message Queuing Protocol}). A comparison of these protocols is given in Table~\ref{tableProtocolsApp}: In addition to the communication model, we compare them based on the category of constrained device on which the protocol can be applied and the header overhead. CoAP and MQTT are clearly a better choice compared to HTTP and AMQP due to their lower packet overhead. CoAP is particularly advantageous for devices with limited resources (class 1) that cannot implement a full IP stack.

\begin{table}
\caption{Comparison of the most utilized application layer protocols in IoT.}
\begin{tabular}{|c|c|c|c|c}
\cline{1-4}
         & Communication model  & Device type & Min. header overhead &  \\ \cline{1-4}
HTTP      & Request-response  &  Class 2  &   8 bytes  &  \\ \cline{1-4}
CoAP   & Request-response   &   Class 1   &  4 bytes  &  \\ \cline{1-4}
MQTT     & Publish-subscribe & Class 2  &  2 bytes  &  \\ \cline{1-4}
AMQP  & Publish-subscribe & Class 2   &    8 bytes     &  \\ \cline{1-4} 
\end{tabular}
\label{tableProtocolsApp}
\end{table}

\subsubsection{Request-response protocols}
\begin{description}
    \item[HTTP] was originally designed for web-based communication. It runs over a TCP (\textit{Transmission Control Protocol}) connection and offers reliable communication, but with a considerable message overhead. Compared to other application layer protocols that are also based on TCP, HTTP is the most verbose and heavyweight, as it adds the largest overhead to TCP packets and overall message size \cite{8088251}. Therefore, HTTP requires high bandwidth, which increases processing latency and energy consumption,  and makes it a less suitable solution for IoT environments. HTTPS, the secure version of HTTP, increases the communication overhead only slightly. In contrast, a secure variant of MQTT significantly degrades the overall communication efficiency of MQTT. The reason for this lies in the differences between the request-response and publish-subscribe communication models and their typical use-cases. The overhead introduced in the request-response model is usually added to larger and less frequent payloads, while a secure connection is usually established once per request. In contrast, the publish-subscribe model implies a persistent connection that is more demanding in terms of maintaining the secure state, and since it is designed for frequent and small messages, the overhead is more significant relative to the payload size.
    \item[CoAP] is a communication protocol that is tailored for constrained devices. It implements the request-response communication model similar to HTTP, but also supports the publish-subscribe communication model. Compared to HTTP, the main difference is that CoAP utilizes UDP instead of TCP at the transport layer. This significantly reduces the communication overhead, but also makes  communication less reliable \cite{8088251}. The authors therefore conclude in \cite{7745303} that the disadvantages include erroneous packet delivery and inability to support complex data types. However, communication with CoAP can be performed using either non-confirmable or confirmable messages: the latter is recommended when higher reliability is required. CoAP performance depends on the network status, and  deteriorates in case of network congestion, leading to retransmission, increased energy consumption, higher latency, packet loss, and reduced throughput. These drawbacks are targeted with new extensions of CoAP, mostly based on different Retransmission Timeout (RTO) strategies that focus on minimizing the number of retransmissions \cite{DONTA2022727}. All these characteristics make CoAP the most suitable option for constrained devices in IoT environments when the request-response communication model is required.
\end{description}

\subsubsection{Publish-subscribe protocols}

\begin{description}
    \item[MQTT] is a lightweight publish-subscribe protocol commonly used in various distributed systems, including IoT solutions. Its basic version runs over TCP connections, but there is also a specialized version for sensor networks named MQTT-SN, which is adapted for constrained sensor devices to run on top of UDP and to utilize reduced payload sizes \cite{DONTA2022727}. The protocol architecture includes a simple broker that receives published messages and forwards them to the connected subscribers. The broker also buffers the messages for further subscribers and offers three different levels of Quality of Service (QoS): at most once, at least once, and exactly once. The levels determine the importance of message delivery, but at the cost of sending additional control messages. The authors in \cite{9247996} state that the advantage of MQTT is its lightweight implementation and energy efficiency, while the main disadvantage is that secured MQTT messages over TLS significantly increase the messaging overhead, while secure communication without TLS is not possible. However, due to its lightweight architecture, energy efficiency and low latency \cite{IQBAL2023109640}, it is probably the most commonly used protocol in IoT implementations today.
    \item [AMQP] is a more advanced protocol than MQTT, offering higher reliability and improved security \cite{9023812}. The difference to MQTT is that the messages are not routed directly from broker topics to consumers. An AMQP broker consists of \textit{Exchanges} and \textit{Queues} that are connected through bindings. There are several types of \textit{Exchanges} which determine broker behavior, while a binding is associated to a routing key which filters out messages for specific queues. Finally, each \textit{Queue} should have one consumer since the messages are lost after they are consumed from it. Due to these characteristics, AMQP has a larger header size fixed to 8 bytes, compared to the MQTT header of 2 bytes. When comparing the performance with the MQTT protocol, the authors in \cite{8914552} state that AMQP offers a higher throughput, but at the cost of a larger packet size and packet loss rate. In \cite{9559032}, the authors state that both protocols achieve similar results when comparing latency and CPU load, although MQTT is slightly more lightweight. 
\end{description}

\subsection{Efficient composition of IoT communication technologies and protocols}

\begin{figure}[htbp]
\centerline{\includegraphics[width=3in]{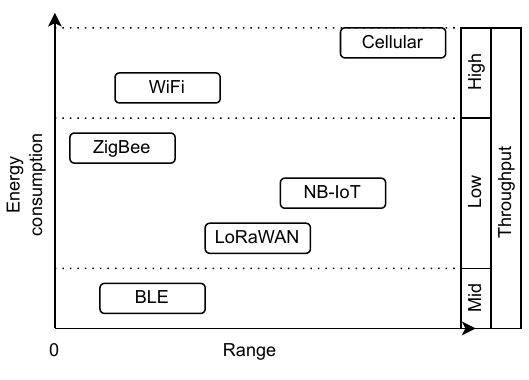}}
\caption{Comparison of the described communication technologies.}
\label{fig:commTech}
\end{figure}

Since there are many technologies and protocols for IoT environments, each with their specific advantages and disadvantages, solution architects have to consider a variety of factors to select the optimal stack of technologies and standards for a given use case. Connecting remote devices to the next layer of the IoT stack is the first step in ensuring efficient data transmission from a physical environment to the gateway nodes. The choice of the right communication technology depends on many factors where the benefits should be weighed against each other. In addition, there are limiting factors that determine whether a particular technology can support a particular use case at all. These primarily include range and throughput, which are usually the starting point when choosing the communication technology for an IoT use case.
Figure \ref{fig:commTech} shows the gradation of the described technologies in terms of range and energy consumption, while they are classified into three groups based on their throughput (low, medium and high). In a simple scenario, a solution architect should choose the most energy-efficient technology offering sufficient range and throughput for the expected environment and generated data. In reality, a scenario can have other priorities that can alter this best practice process of choosing a suitable communication technology. In such cases, a balance must be found between priority requirements and use case-specific factors.

\begin{figure}[htbp]
\centerline{\includegraphics[width=5in]{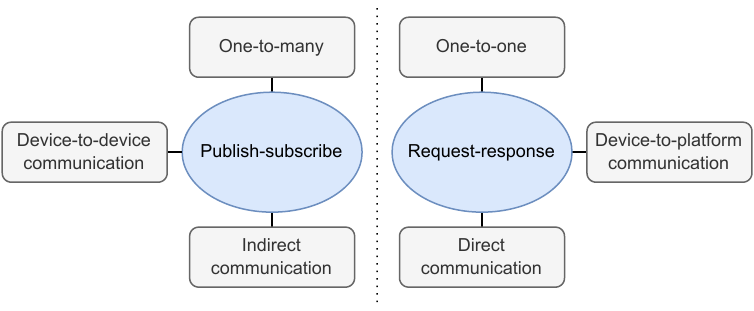}}
\caption{Comparison of the publish-subscribe and request-response communication model for IoT.}
\label{fig:applicationLayer}
\end{figure}

After IoT data reaches a gateway from an end device, it is transmitted to the cloud data centers for further processing or directly towards specific end users. In both cases, it is necessary to ensure a communication architecture that will efficiently deliver the data from the network entry point towards the final destinations. As previously described, there are numerous application layer protocols based on the two major communication models, the publish-subscribe or request-response model. The first step is to decide which of these two models is more suitable for a specific use case (Figure \ref{fig:applicationLayer}). 

There are no best practices for such a decision: The publish-subscribe model offers an efficient architecture that can deliver the data from one publisher to all interested parties without connecting them directly. This loose coupling is beneficial for the IoT-edge-cloud continuum, as participating devices can dynamically enter or leave without affecting the rest of the system. Thus, the publish-subscribe model has the advantage in scenarios where scalability and flexibility are the primary goals. Also, in cases where one-to-many communication is required, while the data does not need to be processed before the delivery to its receivers, it is more advisable to choose the publish-subscribe model.

The request-response model implies a tightly coupled communication, where the data is delivered to a client on request. This type of communication can ensure lower latency because it requires a direct connection between a sender and receiver. Since IoT devices do not usually communicate directly but via an entity such as an IoT platform, it is not recommended to utilize the request-response model for their direct communication. However,  this model is a better choice for communication from a gateway and edge nodes with the IoT platform, or also between FE and NE devices where the data is processed or stored.

After selecting a suitable communication model, the solution architect must select a suitable protocol for the application layer. In the case of the request-response model, CoAP is currently more suitable than HTTP due to its aforementioned advantages. Nevertheless, the HTTP protocol is also widely used since it offers a high degree of interoperability with other web-based systems, as it is the most commonly used protocol for communication over the Internet. In the case of the publish-subscribe model, the choice is limited to MQTT and AMQP. MQTT is widely used because it is simple and lightweight for broadcast communication, but AMQP might be a better choice when higher delivery control and throughput is required.

If the only requirement for an IoT solution is to minimize data traffic within an IP network, the choice should be limited to CoAP and MQTT. CoAP is optimized for constrained devices and its lightweight design ensures minimal protocol overhead, making it the best choice for direct communication with less traffic. However, for more dynamic and scalable environments, MQTT is a more efficient option which ensures reliable publish-subscribe communication with minimal overhead. 

After selecting the appropriate communication technologies for a particular use case, further reduction of network traffic in the IoT-edge-cloud continuum depends on the type of data generated and transmitted in a particular environment and the frequency of message transmission. Therefore, for further discussion on network traffic reduction in the continuum, we must consider the available data reduction strategies which can be used in addition to the available communication protocols and technologies.

\section{Data reduction strategies for the IoT-edge-cloud continuum} \label{sec: dataRed}

Most IoT applications are characterized by a substantial volume of data produced by connected devices, which is largely due to the historical disregard of data transmission efficiency by the majority of existing solutions. However, as IoT adoption continues to grow, the demand for more efficient data transmission is also increasing, as the volume of data generated is doubling annually, according to~\cite{fARHAN}. In this section, we examine the existing data reduction strategies that can help reduce the amount of data traffic sent through communication channels in IoT deployments. They can be grouped into three basic categories:
\begin{enumerate}
    \item Data Compression (DC),
    \item Data Prediction (DP) and
    \item Data Aggregation (DA).
\end{enumerate}

All of these categories aim to reduce traffic to a destination node, but they are based on distinct underlying principles. DC involves encoding the information at the nodes that generate the data and subsequently decoding it at the destination node. This category can be further divided into two primary subcategories: \textit{lossy} and \textit{lossless compression}. 
DP involves the creation of an abstraction of the sensory phenomenon, which is essentially a model describing the generated data. This model is able to predict the values captured by the sensor nodes within a certain acceptable level of error. As the model is applied at a destination node, the need to send sensor data from the source is reduced, provided that the prediction accuracy is satisfactory.
Consequently, DP significantly reduces the need for inter-node communication, which also leads to energy savings.
Data reduction can also be achieved by DA, e.g. by transmitting average values instead of raw values at intermediate edge nodes between a data source and its destination, thereby reducing the amount of data transmitted over the network.

\subsection{Data compression} \label{sssec:compression}
Data compression (DC) methods are applied in many domains; however, this paper specifically examines their application in IoT environments. DC is defined as the process of representing a piece of content while reducing the amount of data required to do so. In the IoT context, the mechanisms of DC reduce the size of information to be transmitted or stored, thereby reducing resource consumption to meet the requirements of the communication and storage infrastructure \cite{NASSRA2023100806}. Several variants of DC methods have been proposed \cite{Lossy_review_iot} which can be classified as follows:
\begin{itemize}
    \item static or dynamic (depending on the variability of the code words over time);
    \item symmetric or asymmetric (in terms of the ratio between compression and decompression execution times);
    \item semantic-dependent or entropy-based (refers to whether or not the data semantics is exploited for better compression);
    \item variable or fixed-rate (related to the data encoding rate);
    \item lossless or lossy (in relation to the decompression results).
\end{itemize}

DC techniques involve two processes: compression and decompression. Lossless algorithms are those that can reconstruct the signal into one that is identical to the original. In contrast, lossy algorithms allow the reconstruction of only an approximation of the original signal \cite{gubitci} to represent the signal with fewer samples than lossless algorithms. In lossy approaches, sensor measurements are usually represented by eliminating redundant samples or by representing them in a compressed form, assuming that redundant samples would not provide additional information for the application \cite{kompresija_2}.  
These methods provide irreversible compression, as it is impossible to recover the original data after compression, while they usually have a lower computational complexity and higher compression results than lossless methods.

The focus of this overview is on lossless or lossy compression classification because sampled IoT sensor data is often used to represent variable signals. Thus, it is crucial to analyze the resulting signal after the compression and decompression steps of DC methods, as it is later used for analysis (e.g., for machine learning, ML) or decision-making (e.g., to trigger alarms or control actuators) in various systems.
The most important differences between lossless and lossy DC methods are outlined in Table \ref{comparison_lossy_lossless}.

\begin{table}[h!]
  \centering
  \caption{Comparison of lossy and lossless compression.}
  \resizebox{\textwidth}{!}{
  \begin{tabular}{p{2.5cm} p{6cm} p{6cm}}
    \toprule
    \textbf{Key} & \textbf{Lossy Compression} & \textbf{Lossless Compression} \\
    \midrule
    Efficiency & More efficient in data size reduction. & Provides moderate data size reduction while preserving data integrity. \\
    \midrule
    Use Cases & Suitable for applications with acceptable data loss (e.g., streaming media). & Ideal for situations where data integrity is critical (e.g., archiving). \\
    \midrule
    Compression Ratios & Achieves higher compression ratios, resulting in smaller file sizes. & Provides lower compression ratios compared to lossy compression. \\
    \midrule
    Reversibility & Irreversible; original data cannot be fully recovered. & Reversible; allows complete restoration of the original data. \\
    \midrule
    Encoding Complexity & Computationally less complex. & May involve more complex encoding processes. \\
    \midrule
    Quality Loss & Degrades the quality of the compressed data due to information loss. & Maintains the original data quality without degradation. \\
    \midrule
    Size & Drastically reduces data size. & Reduces data size, though to a slightly lesser extent than lossy compression. \\
    \midrule
    Decompression Speed & Generally has faster decompression speeds. & Slightly slower decompression speeds due to more intricate encoding. \\
    \midrule
    Sensitivity to Data Types & Can be sensitive to the type of data, leading to variations in quality. & Less sensitive to data types, ensuring data fidelity. \\
    \midrule
    Space vs. Fidelity Trade-off & Prioritizes space savings at the cost of fidelity. & Balances space and data integrity, suitable for long-term storage. \\
    \bottomrule
  \end{tabular}
  }
  \label{comparison_lossy_lossless}
\end{table}

\subsubsection{Lossless compression}
Lossless compression refers to methods where no data is sacrificed during the compression process. These techniques condense data into a smaller file by employing internal abbreviations to mark redundant information. In the IoT context, lossless compression includes a range of algorithms and strategies, including Run-Length Encoding (RLE), Huffman coding, Lempel-Ziv-Welch (LZW) compression, as well as mathematical and statistical approaches. These methods serve to remove redundancy within the data, such as recurring values or patterns, enabling DC while retaining the capability to fully restore the original data if required. 

In a study on compression algorithms~\cite{lossless}, two key aspects were examined: compression efficiency and resource utilization on embedded devices, which is a significant factor for IoT. Adaptive algorithms outperformed static ones, especially on unreliable communication channels, showing the potential for a significant increase in data transmission capacity with minimal time and power consumption. Furthermore, a Two-Tier Data Reduction (TTDR) approach presented in \cite{two_tier} showed that utilizing DC through temporal data correlation at sensor nodes can reduce transmitted datasets by 62\% and lower the device power consumption.
In addition, temporal data correlation enabled Sprintz et al. \cite{sprintz} to achieve high compression with minimal memory and latency overhead for low-power IoT devices, while the authors in \cite{delta} achieved energy savings when dealing with temporally correlated data.

In summary, lossless compression plays a significant role in IoT to optimize data handling and transmission by reducing data size without sacrificing data accuracy, making it a valuable tool for managing high data volumes generated by IoT devices.

\subsubsection{Lossy compression}
Lossy compression methods are a DC approach in which less critical or redundant data is intentionally removed to achieve a significantly higher compression ratio, as described in \cite{lossy}. In contrast to lossless compression, lossy compression intentionally sacrifices the precision of the data in order to significantly reduce the size of the compressed data. In the context of IoT, it proves beneficial for improving data transmission and storage efficiency, especially when limited data loss is acceptable. This helps to save network bandwidth and storage capacity, but should be used with caution as it is not suitable for applications that require complete data accuracy. The key is to find the right balance between compression and acceptable data loss, depending on the specific requirements of IoT applications.

Lossy compression algorithms can be broadly categorized based on the method employed to achieve compression. The four main categories are outlined and compared in Table \ref{evaluation_lossy}.

\begin{table}[h]
  \centering
  \caption{Evaluation of lossy compression strategies.}
  \resizebox{\textwidth}{!}{
  \begin{tabular}{p{3.0cm} p{6.5cm} p{6.5cm}}
    \hline
    \textbf{Approach} & \textbf{Advantages} & \textbf{Limitations} \\
    \midrule
    \textbf{Interpolation} & Lower computational complexity & Lower bandwidth consumption \\
     & Simple implementation & Inadequate representation of the original sample \\
     & Suitable for real-time compression & May not achieve significant compression for highly random data \\
    \midrule
    \textbf{AI Compression Methods} & Adaptability to complex data with high-dimensional features & High computational complexity \\
     & Handling numerous inputs with low error tolerance & Require pre-training \\
     & Improved compression performance through ML & Vulnerability to overfitting \\
    \midrule
    \textbf{Transformation Methods} & Compatibility with various data types and formats & Require data vectors as inputs \\
     & Ability to capture latent data patterns for efficient compression & Need a larger number of measurements \\
    \midrule
    \textbf{Hybrid Methods} & Enhanced compression efficiency through a combination of techniques & High computational complexity \\
     & Efficient data size reduction & Slower response time, in cases where multiple methods are involved \\
    \hline
  \end{tabular}
  }
  \label{evaluation_lossy}
\end{table}

\begin{description}
\item[Interpolation methods] stand out for their lower computational complexity when compared to other techniques. These methods typically employ threshold tolerance parameters to effectively manage compression errors, making them especially suitable for real-time DC applications, such as those involving sensor measurements in the time-domain (i.e. time-series data). 
\item [Hybrid methods] leverage a combination of both lossy and lossless DC techniques. However, it is important to note that hybrid approaches can bring in the disadvantage of higher computational complexity. 
\item [Transformation methods] focus on converting data from one domain to another, which often yields favorable compression results. These methods necessitate data vectors as inputs and require a reasonable number of measurements to attain satisfactory compression outcomes. 
\item [AI methods] tend to entail more intricate implementations and often require a pre-training process. These approaches are particularly well-suited for scenarios characterized by numerous inputs and complex system data.
\end{description}
The above observations are consistent with the findings documented in \cite{Lossy_review_iot}, where a comprehensive analysis of diverse lossy compression methods is reported, with a specific focus on their relevance within IoT environments.

\subsection{Data aggregation}

\textit{Data aggregation} (DA) combines multiple sensor data points while eliminating redundant information. This process is orchestrated by an edge device, such as a gateway or router, which performs the initial data processing and subsequently forwards the information in its processed state instead of the raw, unprocessed data. The importance of this process is particularly evident due to the inherent limitations of sensor devices: They often have limited memory capacity, and their buffers may not be sufficient to hold all incoming data, potentially leading to packet loss. Apart from the data loss, this can put a significant strain on network traffic as nodes have to retransmit the dropped packets, resulting in increased energy consumption on sensor devices and increased latency. To overcome these challenges, DA involves the integration of correlated data at intermediate nodes referred to as \textit{aggregators} which can run on edge nodes. Depending on specific application requirements, an aggregator applies aggregation functions to the received data, e.g. numerical or statistical operations such as minimum, maximum, sum, average, and others. As a result, DA not only reduces the number of data packets and the overall network traffic, but also the latency and power consumption on edge nodes, as elaborated in \cite{aggregation_survey}.

DA protocols define the standard operating procedures to:
\begin{itemize}
    \item aggregate the collected data based on an aggregation function,
    \item handle the exchanges of data and control messages in an IoT environment, and
    \item forward the aggregates to the base station (BS)/edge node/cloud.
\end{itemize}
The primary objectives of DA protocols include the elimination of redundant data transfers from source nodes to higher layers of the IoT architecture, the preservation of data accuracy during aggregation, and the enhancement of the operational lifetime of the entire IoT solution.

There are a number of studies on the application on DA protocols with diverse operational characteristics which take into account: the topology of an IoT environment and its dynamic aspects, as well as security considerations.
A comprehensive exploration of existing methods for DA in IoT was conducted by Pourghebleh et al. \cite{agg_sur}. The authors classify the methods as tree-based, cluster-based and centralized, and asses them based on energy efficiency and latency.  
Compressed sensing, a powerful tool for efficient data acquisition and signal reconstruction, shows promise for IoT DA. However, it often comes at the expense of energy efficiency and recovery fidelity, and it is a challenge to find a balance between a prolonged lifetime of an IoT deployment and resource utilization of constrained devices. Amarlingam et al. \cite{agg_comp} tackled this issue with the Light Weight Compressed Data Aggregation (LWCDA) algorithm. It divides the network of IoT devices into non-overlapping clusters, reducing data transmissions within each cluster. When the number of IoT devices increases and data redundancy becomes prevalent, the key challenge lies in aggregating non-redundant data.
Idrees et al. \cite{agg_mech} devised a two-stage DA mechanism named IDiCoEK, using a divide-and-conquer algorithm to eliminate redundant data. Their approach employs an extended K-Means algorithm to filter out redundancies and determine optimal representative data for aggregation. Moreover, the use of data abstraction techniques can be helpful. The approach proposed in~\cite{added} supports fault diagnosis and prediction in complex socio-technical systems, and emphasizes the interaction between technical and social components in various applications.

The IoT-edge-cloud continuum is an optimal computing environment which can offer localized data access, reduce transmission delays and enable decentralized DA, resulting in more efficient resource management and energy conservation for IoT and edge devices~\cite{2020edge}.

\subsection{Data prediction}

In the context of IoT, DP employs advanced analytics and ML algorithms to forecast future data trends based on historical data patterns. This predictive capability enables IoT devices and networks to proactively optimize data transmission and reception patterns, thereby improving inefficiencies in data traffic management \cite{OLIVEIRA2024101153}.

In DP methods, a model is created that represents a captured phenomenon and allows requested values to be retrieved based on the predictions of the model rather than the directly sensed data. Many prediction methods have been proposed in literature as effective tools to reduce data collection and transmission rates in IoT. These techniques can be broadly categorized into two main approaches based on the location of the model within an IoT environment~\cite{dias2016survey}: 1) single prediction and 2) dual prediction.
\begin{description}
    \item[Single prediction:] In the \textit{single prediction} approach, predictions are made at a single point in the network, such as the sensor node or edge gateway. The method outlined in \cite{hussein2022distributed} involves implementing a prediction algorithm at the sensor node level. In this approach, the data readings gathered by a sensor node in the initial period are utilized to predict the sensor readings for the subsequent period. These sensor readings from the first period are always sent to the gateway after compression. In the following period, the similarity between the predicted data and the new sensor readings is then computed. If the similarity score exceeds a predefined threshold, the new data is not transmitted to the gateway. An alternative approach to single prediction is to alternately schedule data transmission and DP \cite{10143415}. During the DP period, the collection of actual measurements from sensor nodes is omitted. The authors employed a deep learning technique to develop a prediction model deployed on gateways that allows prediction of sensor readings. 
    \item[Dual prediction:] In \textit{dual prediction} systems, the prediction models are utilized both at the IoT nodes and in the IoT-edge-cloud continuum. An IoT node continuously compares the current observation with the predicted value. If the measured value falls within the predefined error range, the model remains valid. However, if the measured value deviates beyond this threshold, the source node can transmit the sampled data and potentially initiate a model update procedure. The design and characteristics of a DP method depend essentially on the specifics of the model's construction. In~\cite{dual_prediction}, the authors propose a reliable dual prediction data reduction approach for Wireless Sensor Networks (WSNs). This approach operates through two phases to achieve data reduction. The first phase aims to decrease the number of transmissions between the sensor node and the sink node, thereby minimizing energy consumption. It also detects faulty data at the sensor nodes and discards them. The discarded faulty data at the sensor nodes are replaced by estimated values at the sink node to maintain data reliability. The Dual Prediction Process (DPP) runs at the sink node or base station and works synchronously with the sensor nodes. This phase is responsible for predicting the non-transmitted data based on the Kalman filter. The simulation results demonstrate that the proposed approach is efficient and effective in reducing data, ensuring data reliability, and minimizing energy consumption. In another study~\cite{PLACZEK2024225}, the authors present a method to reduce data transmissions in sensor networks when selecting target nodes for monitoring tasks. Their approach empowers an IoT gateway to decide when sensor nodes should report their data readings. By utilizing a prediction algorithm that employs an instance-based learning technique to assess potential future change rates of sensor readings, the gateway determines the need to select a new target node and transmits data only when essential for accurate target node selection. Experiments were conducted in a WSN with mobile devices, where the target node was the device closest to a specified location. The results show that the proposed method can significantly reduce the transmitted data and accurately select the target node.
\end{description}

The prediction algorithms can be broadly categorized into three classes:
\begin{itemize}
    \item \textbf{Stochastic approaches}: Rely on statistical properties and probabilities to characterize the phenomenon. They often involve mapping data to a random process described by a probability density function (pdf) and predicting future data by combining computed pdfs with observed samples;
    \item \textbf{Algorithmic approaches}: Utilize heuristic or state transition models to describe the phenomenon being analyzed. These approaches do not rely solely on statistical properties but instead focus on building and updating models based on predefined algorithms or procedures; and
    \item \textbf{Time series forecasting}: Involves using historical values from a time series obtained through periodic sampling to predict future values within the same series.
\end{itemize}

Since time series data streams are frequently generated by devices in IoT environments, time series forecasting is a prevalent DP method in literature. In addition, stochastic and algorithmic approaches can be computationally intensive and may require significant processing power and memory. Since IoT and edge devices have limited resources, implementing complex stochastic models may not be practical and thus in this paper we further investigate time series forecasting. 

Time series forecasting uses a series of historical values obtained by periodic sampling to predict a future value in the same series. The main difference to other statistical or probabilistic approaches is that time series analysis explicitly considers the internal structure of the data. In general, a time series can be represented as a combination of a pattern and a random error, where the pattern is in turn characterized by its trend, i.e., its long-term change, and its seasonality, i.e., its periodic variation. Once the pattern is fully characterized, the resulting model can be used to predict future values in the time series. ML architectures used in literature for time series forecasting include~\cite{review_forecast}: 
LSTM (Long-Short Term Memory) because of their high prediction accuracy and ability to automatically process input sequences;
 CNN (Convolutional Neural Networks), which are mainly used for human activity recognition; hybrid architectures with a convolutional layer for data preprocessing;  
RNN (Recurrent Neural Networks) for data fusion from different sensors and their subsequent classification; and
stacked LSTM autoencoders, which extract variables from time series without manual data preprocessing. 
For example, the study reported in~\cite{jarwan2019data} demonstrated that the effectiveness of reducing data transmission in an IoT system depends largely on the specific prediction algorithm used. The authors conducted a comparison of different prediction algorithms based on their accuracy, delay, and the percentage of reduction in transmission. Additionally, neural networks and LSTM networks have been proposed for dual prediction schemes with online model training.

Another approach that can be taken into consideration is using DP together with model compression methods. Model compression techniques, such as pruning, quantization, and knowledge distillation, are essential for optimizing neural networks in resource-constrained IoT environments. These methods help reduce the model size and computational requirements, making it feasible to deploy complex prediction algorithms on IoT devices \cite{model_compression}.
\begin{itemize}
    \item \textit{Pruning} involves removing redundant or less significant weights and neurons from the neural network, thereby reducing its size to improve inference speed without significantly affecting performance \cite{cheng2023surveydeepneuralnetwork}.
    \item \textit{Quantization} reduces the precision of weights and activations from floating-point numbers to lower-bit representations (e.g., 8-bit integers). This not only decreases the model size but also enhances inference speed, especially on hardware that supports lower-precision arithmetic \cite{Hernandez2024}.
    \item In the \textit{knowledge distillation} approach, a smaller, simpler model (the student) is trained to replicate the behavior of a larger, more complex model (the teacher). The student model learns to mimic the teacher's predictions, allowing it to achieve similar performance with significantly fewer parameters \cite{knowledge}.
\end{itemize}
These techniques are key to ensure that predictive models can run efficiently on IoT devices, which often have limited processing power and memory. While our work focuses on data prediction to minimize data transmission, the inclusion of model compression methods can further enhance the feasibility and efficiency of deploying predictive models in real-world IoT applications.

\subsection{Choosing the right data reduction technique}
Selecting the most suitable technique for data reduction in the IoT-edge-cloud continuum requires careful consideration of its efficiency and accuracy by solution architects at each layer of the continuum. These techniques play a critical role in optimizing the efficiency of data transmission, storage, and processing within IoT systems \cite{SADRI2022100629}. The primary objective is to minimize the amount of data transferred in the continuum while preserving important information for analysis and decision-making. 

The choice of the appropriate technique depends on several key factors:
\begin{itemize}
    \item \textbf{Data type and required accuracy}: The type of data being collected and the accuracy required for the analysis significantly influence the choice.  For example, applications that focus on general trends may utilize DA effectively, while applications that require high accuracy of measurements may favor minimal lossless compression techniques.
    \item \textbf{Available resources}: The processing power and memory constraints of the deployed IoT devices play a crucial role. Resource-constrained devices at the edge necessitate simpler techniques such as data filtering or lightweight compression, while gateways or cloud servers can handle more sophisticated methods such as ML or neural networks.
    \item \textbf{Trade-off between efficiency and accuracy}: It is important to find a balance between the efficiency of data reduction while preserving data accuracy. Techniques such as DC can significantly reduce transmission volume, but may result in loss of detail, especially with lossy algorithms. Conversely, data filtering transmits only relevant data, but may discard potentially valuable information.
\end{itemize}

\subsubsection{Comparison of data reduction techniques}

\begin{figure}[ht]
    \centering
    \includegraphics[width=0.8\textwidth]{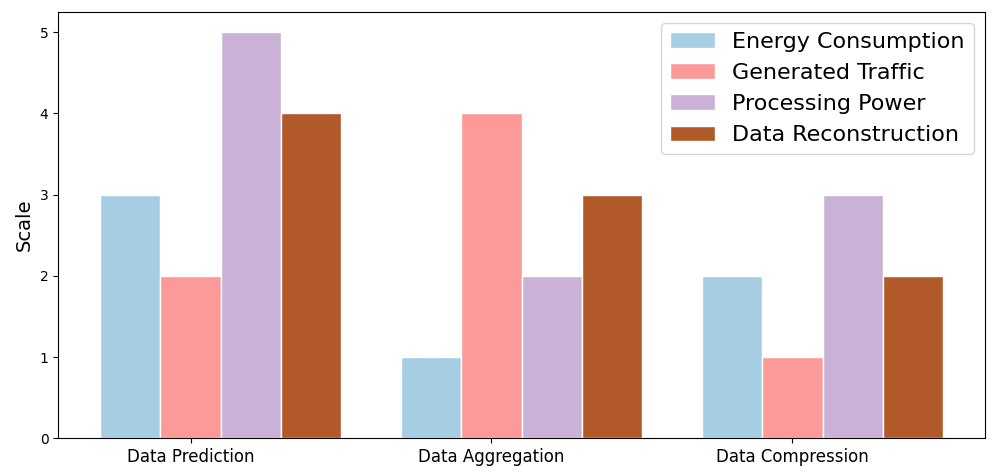}
    \caption{Comparison of data reduction methods for IoT environments.}
    \label{fig:comparison}
\end{figure}

Unlike communication protocols, a direct comparison of data reduction techniques without a given usage context is challenging due to the inherent variability of these methods. DP algorithms, for example, leverage diverse approaches such as ML or neural networks, each with distinct computational demands and characteristics. Additionally, the size of the deployed model itself significantly impacts the overall efficiency. Therefore, the effectiveness of each method depends heavily on the specific application, dataset properties, and the underlying IoT system's resource constraints. To provide a clearer basis for comparison, Figure~\ref{fig:comparison} offers a comparative analysis of the presented techniques based on four key criteria crucial for the data reduction in the IoT-edge-cloud continuum, where we rate each criteria on a scale from 0 (the lowest) to 5 (the highest):

\begin{enumerate}
    \item \textbf{Energy consumption}: The technique's impact on device battery lifetime and overall system energy usage;
    \item \textbf{The volume of generated traffic}: The amount of data transmitted when applying the technique;
    \item \textbf{Required processing power}: The computing resources required by the device to implement the technique; and
    \item \textbf{Accuracy of data reconstruction}: The ability to recover the original data from the reduced representation.
\end{enumerate}

Generally, DA offers a good balance between efficiency and complexity. It combines multiple data points into a single, representative value, reducing data transfer but potentially sacrificing some detail. However, it proves to be an energy-efficient approach. In contrast,
DC is highly effective in reducing data traffic with moderate energy consumption, making it a compelling choice for minimizing data transmission in IoT environments. However, the potential loss of detail, especially with lossy algorithms, needs to be considered. DP is considered valuable for forecasting future values based on historical trends, but prediction methods require more processing power and energy compared to other techniques.
Optimal selection  often involves a combination of these techniques, tailored to achieve the best outcomes for a specific IoT use case.  Furthermore, the placement of these methods within the IoT architecture has a significant impact on their effectiveness, as explored in detail in the following subsection.

In summary, each data reduction technique offers distinct advantages and limitations. Careful evaluation of the above factors, as well as their strategic placement within the IoT architecture, is critical to select the most suitable approach to achieve the desired balance between efficiency and accuracy in an IoT solution.

\subsubsection{Placement of data reduction techniques within the IoT-edge-cloud continuum}

The placement of data reduction techniques within the IoT-edge-cloud continuum, as illustrated in Figure~\ref{fig:edgeStack}, is another key design consideration. Solution architects must strategically select IoT and edge devices which generate, transmit and process data streams that could host a data reduction technique which offers the greatest overall effectiveness. Several factors influence the placement of these techniques. Firstly, the \textit{specific requirements of an application and the capabilities of the devices} involved need to be considered. This involves balancing the desired level of data reduction with the processing capacity of the deployed devices. For instance, resource-constrained sensor nodes located closer to a data source are better suited for simpler techniques, while devices with higher processing power, such as gateways or cloud servers, can handle more complex algorithms. Secondly, a \textit{desired level of data reduction and the communication protocols used} within the environment also affect placement decisions. A higher level of data reduction may require more processing power, potentially shifting placement closer to higher capacity devices at the near edge or even in the cloud. Moreover, the efficiency of communication protocols within the architecture plays a significant role. 
By carefully weighing these considerations, architects can strategically position data reduction techniques, thereby optimizing resource utilization and achieving the desired level of data reduction within the constraints of the IoT application.

Building on the importance of strategically placing data reduction techniques within the IoT-edge-cloud continuum, a more detailed exploration of specific approaches suitable for different layers can be undertaken. Hereafter we examine the techniques employed on resource-constrained devices at the device and edge layer.

\paragraph{Data reduction at device layer.} Data processing on end devices is limited by the processing performance, memory and available energy of the microcontroller. In many cases, a microcontroller is connected to the network via wireless technology and is powered by a battery. In this case, in addition to processing performance and memory size, energy consumption is also important, as it is consumed for generating sensor readings, data processing and communication. Since devices may be located in places where battery replacement is impractical or costly, energy efficiency is crucial for many application domains, e.g. environmental monitoring or smart agriculture. Limited memory and processing power enable IoT devices to perform only simple data analysis. DC is a suitable technique for  data reduction on such devices \cite{compression_iot} which removes redundant data before transmission to the gateway, thus maintaining data quality while reducing volume. If the devices are categorized as Class 2, or occasionally Class 1, more demanding reduction methods can be used, such as DP with a more lightweight algorithm which employs  an ML inference model that can run on the available hardware.

\paragraph{Data reduction at edge layer.} Edge devices, such as efficient data servers, routers, or embedded systems, receive data from IoT devices  and vary widely in hardware and operating systems. Gateways, which are often single-board computers like Raspberry Pi, have more processing power (typically an ARM processor) and large amounts of memory (up to several GB), allowing for more complex data processing than at the device layer. In addition, gateway nodes can collect and store data from multiple endpoints, analyze it, and send commands rapidly either upwards toward the cloud or backwards to an actuating device, making DA techniques appropriate for gateways. 
DP techniques can also be implemented on edge nodes, especially on near edge devices, as they have higher processing power compared to IoT devices \cite{gateway}. Neural networks can  be deployed on gateways to handle missing or noisy data \cite{missing}. Another approach involves deploying an ML model on the edge device, where the model predicts values based on sensor data and compares them to sensed values. If they align well within a predefined threshold, the data is considered accurate, avoiding transmission over the network. In the cloud, a similar model instance is used which calculates the value, instead of receiving it from an IoT devices. This technique can greatly reduce the overall data traffic towards the cloud, but it relies on accurate prediction models. 
In cases when all sensed data needs to be sent to the cloud for further analysis or storage, compression methods can be applied to reduce the packet size, and thus save bandwidth and storage space. In scenarios with multiple devices measuring similar data where aggregation occurs at a higher layer, data merging or joining is effective to reduce the overall transmitted data. This multi-layered approach optimizes resource utilization, improves efficiency, and accelerates decision-making in the IoT-edge-cloud continuum.
To illustrate a specific edge computing scenario, consider a smart agriculture application where various sensors monitor soil moisture, temperature, and humidity levels across a large farm. An edge gateway, such as a Raspberry Pi, collects this data from numerous sensor nodes distributed throughout the farm. Given the redundancy in data due to the proximity of sensors, the gateway employs a data aggregation technique to merge readings from sensors that show similar values. This aggregation reduces the amount of data sent to the cloud by summarizing the sensor data into a single representative value for each parameter within a specific area. Additionally, the gateway uses a lossy compression algorithm to further compress the aggregated data before transmission, striking a balance between data accuracy and transmission efficiency. This approach minimizes the data traffic to the cloud, conserves bandwidth, and ensures that essential information is still available for further analysis and decision-making in the cloud.

\begin{figure}[ht]
    \centering
\captionsetup{justification=centering}
    \includegraphics[width=0.95\linewidth]{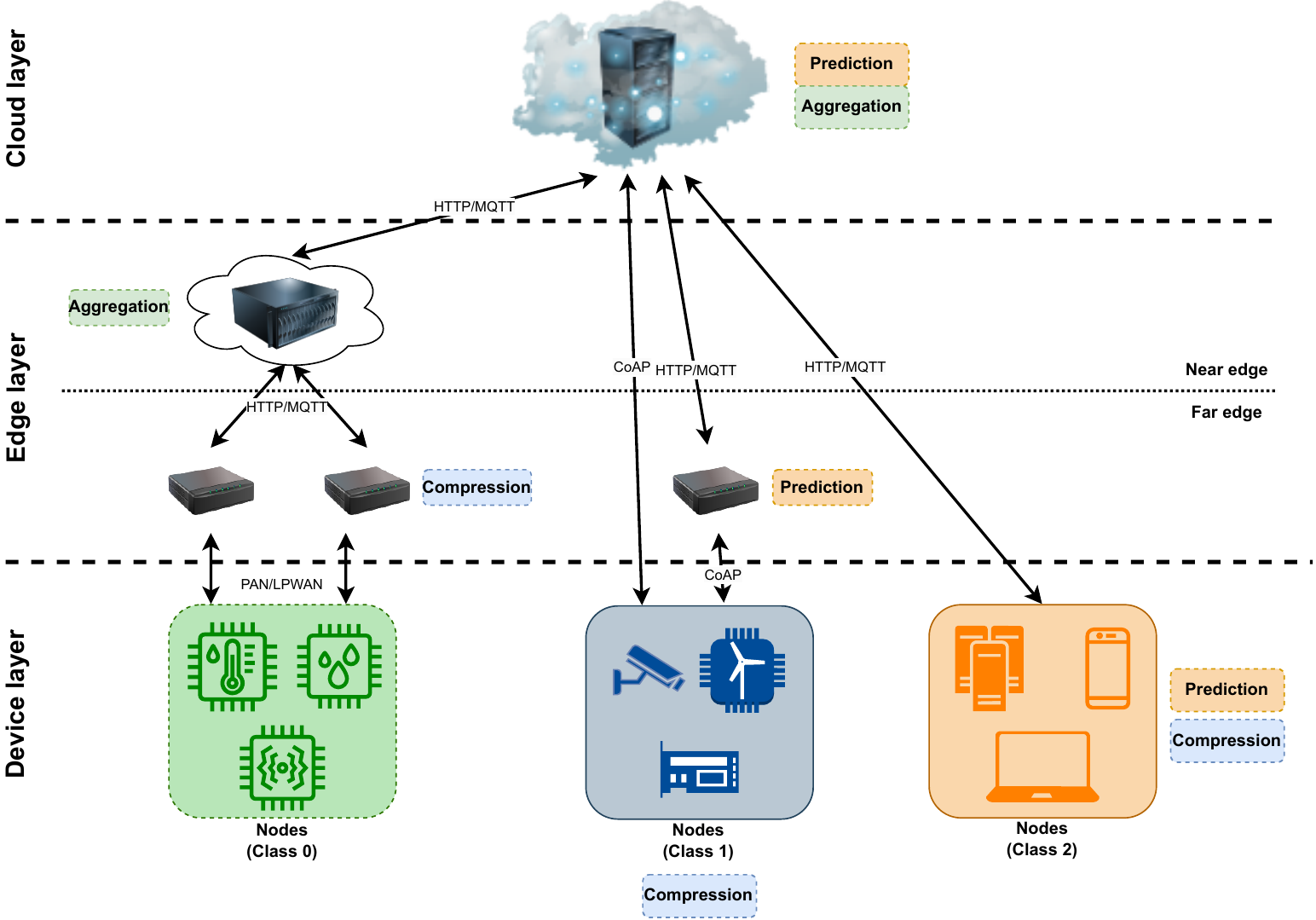}
    \caption{Distribution of data reduction techniques across different layers of the IoT architecture}
    \label{fig:reduction_arh}
\end{figure}

Figure \ref{fig:reduction_arh} showcases a potential positioning of the discussed data reduction strategies across the three layers of the IoT architecture. Within each layer, distinct techniques such as DC, DP, and DA are applied to streamline data processing and transmission.
The application of these techniques varies depending on the characteristics of the devices and the specific requirements of the IoT ecosystem. For instance, in the IoT device layer, devices classified as Class 0 may have limited capabilities, making traditional data reduction methods impractical. However, by transitioning to the far edge layer, efficient DC techniques can be employed to optimize data transmission. Moreover, at the near edge layer, DA plays a pivotal role in balancing efficiency and complexity while preserving essential data details.

Devices classified as Class 1 typically possess sufficient resources to run compression algorithms effectively. In such cases, deploying predictive models at the near edge layer can enhance data processing efficiency by transmitting only significant deviations from predetermined thresholds. In contrast, devices falling under Class 2 have higher capabilities and can employ prediction techniques without the need for intermediate aggregation steps. This approach not only optimizes the use of resources, but also accelerates decision-making processes within the IoT network. Moreover, due to the versatility of DP techniques, specific instances of predictive models can also be seamlessly integrated into the cloud layer. This can ensure that although the raw or compressed data is not transmitted to the cloud, the model makes it available in the cloud for subsequent analysis and aggregation.

In summary, the selection and implementation of data reduction techniques within the IoT architecture are intricately linked to specific use case scenarios, device capabilities, and network characteristics. A strategic deployment of these techniques across different layers and devices can lead to an optimization of IoT-edge-cloud resource utilization, and enhancement of overall system performance.


\section{Research directions and open challenges for efficient data transmission strategies in the IoT-edge-cloud continuum}
\label{sec:challenges}

Choosing the appropriate communication technology, as described in Section \ref{sec: transTech}, is a prerequisite for optimal data transmission in the IoT-edge-cloud continuum, while data reduction strategies presented in Section \ref{sec: dataRed} facilitate information exchange with significantly less data traffic. 
In this section, we outline the emerging concepts that have the potential to further enhance IoT communication by reducing the associated network traffic: cross-layer optimization and edge-AI techniques for data reduction.

\subsection{Cross-layer optimization}

Traditional communication network design follows the Open Systems Interconnection (OSI) model, which separates protocol functionalities into distinct layers. Unfortunately, the technologies within each layer are often developed in isolation without considering their impact on other layers. This lack of attention to cross-layer interactions~\cite{fi16010029} is known to have a significant impact on end-to-end (E2E) delay and reliability.

To overcome these limitations, cross-layer design promotes collaboration between layers for improved network performance. It falls into two categories~\cite{cl}: loosely coupled or tightly coupled cross-layer design, where both are based on establishing cross-layer interactions.

In a \textit{loosely coupled cross-layer design}, the focus is on optimizing a specific protocol layer without direct interaction with other layers. However, disconnected layers share information through interfaces or by using a common database. Thus, information from other layers is taken into account to improve the performance of the targeted layer. For example, the physical layer can pass channel quality information to the TCP layer so that it can differentiate between packet loss and congestion to enable more effective congestion control.

In contrast, a \textit{tightly coupled cross-layer design} goes beyond simple information exchange between the layers. In this approach, layer separation disappears and the algorithms of the different layers are optimized together as part of a single optimization problem. While a tightly-coupled cross-layer design offers superior performance improvements, the advantage of a loosely-coupled cross-layer design is that it maintains transparency between the protocol layers instead of giving it up completely.

Cross-layer design has long been present as a possible approach towards enhancing communication performance through better resource allocation \cite{articleResAllocation}, power control \cite{7986381} and congestion control \cite{4205060}. 
However, this design violates the highly interconnected composition of layers in the OSI model. Each modification of such a model creates many unforeseen consequences that often reduce the efficiency of the entire system. While its development complexity limits widespread adoption, research demonstrates that cross-layer design has potential for optimization in various communication scenarios, particularly for specific Quality-of-Service (QoS) parameters~\cite{10.1007/978-981-19-2719-5_62,HASAN2022101099}. Cross-layer solutions are often targeted and application-specific to improve a specific communication parameter. 

The cross-layer approach was often used in use cases related to WSNs because of the constraints of participating sensor devices \cite{parween2020review}. One popular example that showcases the benefits of OSI layer adaptation in this context is the 6LoWPAN protocol (developed by the 6LoWPAN IETF group) which implies the optimization of IPv6 packets from the network layer for their transmission over low-power wireless networks operating in lower layers, typically based on IEEE 802.15.4. Thus, cross-layer optimization could have a significant role in reducing the overall data traffic in IoT, especially at the edge where participating devices often cannot even support the full IP stack, as previously described.

\subsection{Edge AI techniques for IoT data reduction}
The growth of resource-rich device, e.g with graphics processing units (GPUs)/ neural processing units (NPUs) or field programmable gate arrays (FPGA), at the network edge offers an opportunity to distribute processing tasks from the cloud to devices that are closer to the targeted local environments. 
The execution of data processing closer to the data sources in IoT offers a wide range of opportunities to optimize the overall system performance and achieve benefits such as reduced latency, lower bandwidth, and improved security. This section explores how Edge AI complements data reduction methods presented in Section~\ref{sec: dataRed} by enabling intelligent local data processing and real-time analysis. Various techniques are applied at the edge, and a comprehensive survey of AI-related algorithms suitable for edge applications is presented in~\cite{s23031639}.

Edge AI executes machine learning (ML) algorithms on IoT and edge devices, analyzing data relevance before transmission to the cloud for in-depth analysis. By decentralizing AI processing from the cloud to edge devices, Edge AI enables real-time data analysis, reduced latency and improved privacy protection. This paradigm shift is fundamentally changing the landscape of AI applications and paving the way for a new era of intelligent and responsive systems. This decentralization from cloud-based AI offers several advantages~\cite{SINGH202371}:
\begin{description}
    \item \textbf{Reduced latency}: Edge devices offload processing workload from the cloud platforms, thus leading to a significant reduction in end-to-end latency while also freeing cloud resources for more complex tasks. Especially, with low-latency or real-time applications, the ability to carry ML/AI workloads at edge/fog nodes closer to controlling components is vital because the cloud computing infrastructure has its physical limits in terms of communication delay guarantee. For example, for autonomous vehicles (AV), milliseconds of delay can be critical. Processing sensor data locally ensures immediate responses to changing driving conditions \cite{vehicles}; hence, AVs always have on-board edge computing units such as Nvidia Jetson Orion (6-12 Arm Cortex-A78AEcores and up to 2048 GPU cores) \footnote{\url{https://www.nvidia.com/en-us/autonomous-machines/embedded-systems/jetson-orin/}} and NXP S32G2 Processors~\footnote{\url{https://www.nxp.com/products/processors-and-microcontrollers/s32-automotive-platform/s32g-vehicle-network-processors/s32g2-processors-for-vehicle-networking:S32G2}}(two Arm Cotex-A53 and one Cotex-M7). These edge computing capabilities allow AVs to move safety-critical features such as sensor fusion and driving assistant to in-vehicle network and rely on the cloud only for other features that do not require deterministic delay such as fleet-management and maintenance analytics. Similarly, edge-driven processing and communication is a research direction for moving low-latency workloads such as SLAM and sensor fusions away from the cloud to edge/fog nodes, e.g. FogROS~\cite{DBLP:conf/icra/IchnowskiCDADVJZLXBKSGG23} and Edge Robotics~\cite{DBLP:journals/iotj/HuangZCLZY22} or Swarm-SLAM/Map~\cite{Jingao:2022,lajoieSwarmSLAM}.  
    \item \textbf{Efficient data transmission}: Edge AI can significantly contribute to reducing the required bandwidth by processing, analyzing, and storing data locally. This localized approach means less information needs to be transferred to the cloud, resulting in a streamlined data flow. For instance, in a smart manufacturing setup, edge devices can analyze production line data in real-time to detect anomalies, sending only summary reports to the cloud \cite{SOORI2023192}. Similarly, in various applications where data volumes require high throughput such as video analytics, traffic monitoring and fleet managements when sending all data to the cloud is prohibitively expensive, various AI techniques need to be combined at the edge to reduce the communication bandwidth or cost.
    \item \textbf{Improved security and privacy}: Edge AI contributes to increased security by limiting the amount of data sent to centralized cloud storage. The vulnerability associated with all data residing in a single cloud storage is mitigated as some data is processed and stored within an edge network. When data needs to be transferred to the cloud, Edge AI also filters out redundant, superfluous and unnecessary information and only sends the important data. This is particularly relevant in healthcare applications where patient data privacy is paramount \cite{electronics12041027}. Furthermore, with the introduction GDPR, moving data to the cloud or outside of geographical locations is not an option, Edge AI can help to pre-process data and then send only the intermediate processing states that can be further processed in privacy-preservation manners.
    \item \textbf{Scalability and flexibility}: The ability to seamlessly extend and customize the system is facilitated by distributed processing and analysis at the edge. This inherent scalability ensures that the AI system can grow effortlessly to meet evolving requirements, while the flexibility of Edge AI enables adaptable functionality for different applications and environments.
\end{description}

Edge AI uses various technologies. These include edge orchestration frameworks~\cite{cilicPerf}, ML frameworks, neural network architecture, model compression, federated learning~\cite{ZHANG2021106775}, and edge-to-cloud integration. The implementation of Edge AI software requires efficient hardware that is able to process data from various devices or sensors while consuming as little power as possible. The challenge is to achieve high algorithm accuracy without compromising energy efficiency. The advancement of hardware options such as central processing units (CPUs), GPUs, NPUs, application-specific integrated circuits (ASICs) and system-on-chip accelerators (SoCs) has made Edge AI feasible \cite{8951147}. To support the development and deployment of Edge AI models, ML software frameworks like TensorFlow Lite, PyTorch Mobile, and ONNX Runtime provide valuable tools. These frameworks come equipped with features such as model optimization and deployment tools, optimizing resource utilization and performance on edge devices \cite{liu2022bringing}.

Model compression \cite{li2023model} is a technique that plays an important role in this ecosystem to reduce the size of ML models to fit the resources of edge devices so that inference is performed close to data sources. By eliminating redundant information, it enables the execution of models on nodes with limited resources.

Federated learning (FL) offers a methodology that supports the distributed training of ML models on a federation of participating devices (FL clients) on local data without the need to share raw data with the cloud \cite{ZHANG2021106775}. This approach not only enhances privacy, but also reduces the associated network traffic, since data remains on FL clients, while model updates are sent to a central aggregator in the cloud. The collected models are aggregated and returned back to the FL clients that start a new training round with the updated global model. This iterative process continues until the global model converges. FL saves network bandwidth as the raw data does not leave the FL clients, but only in case when the shared ML models are not large. In addition, the computational and storage costs of the central server are significantly reduced as the training task is split across different devices.

The edge-to-cloud integration creates hybrid solutions that leverage the strengths of both edge and cloud computing. It promotes real-time processing on edge devices, while data synchronization ensures efficient transfer and updating of edge-collected data in the cloud. Simultaneously, task allocation and offloading distributes computing tasks between edge devices and the cloud based on resource availability, latency requirements, and workload characteristics to conserve energy, reduce the associated network traffic and facilitate system scalability. In particular, hierarchical FL (HFL)~\cite{liu_hiear} has recently emerged as an answer to the FL performance problems caused by an increased communication overhead of FL in IoT environments with numerous participating devices and frequent model updates. Furthermore, it is often challenging to ensure model consistency in distributed IoT environments as they imply a high level of device and data heterogeneity. Thus, the edge layer in HFL  the aggregation function for a cluster of FL clients to synchronize and aggregate their local models. The models aggregated at the edge layer are then merged by a global aggregator residing in the cloud. This approach is expected to reduce the overall network traffic, energy consumption and training time compared to traditional FL due to frequent local aggregations which are performed at the edge layer, while global aggregations involving cloud resources are less frequent.


\subsection{Open challenges and trends}
The following are potential open issues and emerging trends for the future of Edge AI:
\begin{description}
    \item \textbf{Data loss}: Implementing edge AI requires careful planning to avoid data loss. Although many edge devices discard irrelevant data properly, there is a risk of losing relevant data, which affects subsequent analyzes. For instance, important sensor data might be discarded if it is not properly categorized, affecting predictive maintenance systems in industrial applications. This challenge can be overcome with generative models which can reconstruct the missing features or signals.
    \item \textbf{Limited computing power}: While effective, edge AI lacks the computing power \cite{s23031279} of cloud-based AI, restricting certain tasks to edge devices. Cloud computing remains essential for large models, while edge devices excel in on-device inference and modest transfer of learning tasks. Implementing edge AI requires balancing high algorithm accuracy with low power consumption. Recent innovations in hardware have made this balance achievable. Prominent companies like NVIDIA, Intel, and Qualcomm are at the forefront of advancing AI capabilities at the edge.
    \item \textbf{Hardware variations}: Diverse edge devices introduce variability in performance, latency, and power efficiency, complicating system reliability. Newer hardware like Nvidia’s Jetson Orin and Intel’s Myriad X offer improved AI capabilities, but hardware differences—whether using GPUs, NPUs, or TPUs—require adaptable models and flexible software to maintain consistency.
    \item \textbf{ML model variations}: Different hardware requires different inference models which match the available resources. New challenges arise since such models offer different accuracy and inference latency. Ensuring consistency across various edge devices in a large-scale deployment, e.g. in agricultural or smart city IoT systems, can be challenging since we need to take into account the performance of ML models in addition to infrastructure-related parameters such as energy consumption and network traffic.
    \item \textbf{Privacy and security}: Ensuring privacy and security in processing data on local devices will be a critical consideration for edge AI. Techniques such as FL, which enable local model training without exchanging raw data, will become increasingly important. For instance, in healthcare, FL can allow hospitals to collaboratively train models without sharing sensitive patient data \cite{flhealth}. Communication-efficiency in FL is still an active research topic, hence, the underlying aspects of protocols and data reductions in Section~\ref{sec: transTech} and Section~\ref{sec: dataRed} are vital elements to effectively deploy FL in the edge devices as discussed and analyzed in~\cite{DBLP:journals/corr/abs-2305-19831,DBLP:journals/corr/abs-2405-20431}
    \item[Energy efficiency] remains a constant concern with edge AI. Developments in this area include the work by Levisse et al., who introduced functionality-enhanced memories \cite{levisse}. Additionally, Liu et al. proposed hybrid parallelism, which enhances the efficiency of hierarchical training for AI models in Edge AI scenarios \cite{9094236}.

\end{description}

\section{Conclusion}
\label{sec:conclusion}

The IoT-edge-cloud continuum is a hierarchical and layered computing architecture that integrates IoT devices, edge nodes with heterogeneous computing resources, and the cloud to form a unified environment for deployment of modern IoT solutions which are responsive, efficient, secure and scalable. However, this environment introduces new challenges due to its complexity and the resource limitations of edge devices compared to powerful cloud resources. In particular, efficient data transmission within this complex ecosystem presents a challenge due to the utilization of heterogeneous hardware, wireless technologies and application-layer protocols. Data reduction strategies are a viable approach to facilitate efficient data transmission in the IoT-edge-cloud continuum, but a holistic view is needed to identify promising techniques and their potential position within the continuum in line with the limitations of the environment. 

This paper provides such a holistic view by analyzing the communication technologies and protocols, data reduction techniques and their combined use and placement with the IoT-edge-cloud continuum.

We first compared the wireless communication technologies for the IoT-edge-cloud continuum based on the range, power consumption, and throughput to conclude that the most energy-efficient protocol which meets the range and data rate requirements of a specific IoT use case should be chosen for communication between IoT devices and the gateway. For communication between the gateway or edge device and the cloud, CoAP, which implements the low-latency request-response model, and MQTT, which offers scalable and flexible one-to-many interaction, are ideal for minimizing traffic within the continuum. Further optimization strategies can be explored based on data types, transmission frequency, and the overall IoT-edge-cloud environment.

Next, we explored the following data reduction strategies applicable at the IoT device, gateway and edge level: DC, DP and DA. They were compared based on  energy consumption, traffic volume, processing power and accuracy. An optimal strategy for data reduction involves a combination of these techniques, tailored to the specific use case with careful planning of their placement within the continuum. DC is commonly deployed on the IoT device and far edge nodes, and proves effective in minimizing data transmission but may involve some data loss with lossy algorithms. DA provides a good balance between efficiency and complexity by reducing data transmission while retaining some detail. DA solutions are deployed at the remote or near edge nodes. DP is valuable for forecasting future values, but requires more processing power and energy. It would be beneficial to deploy DP techniques on IoT devices, but due to limited resources far edge nodes are a better choice.

Finally, we explored cross-layer optimization and Edge AI techniques, which are promising and emerging concepts with potential to further reduce the communication overhead in the IoT-edge-cloud continuum. Cross-layer optimization promotes collaboration between layers for improved network performance and shows potential to optimize various Quality-of-Service (QoS) aspects like resource allocation, power control, and congestion control. However, the complexity of development and potential to cause unintended far-reaching consequences limit its widespread adoption. Edge AI shows further potential to reduce data traffic in the continuum by placing ML models across the continuum to perform real-time, intelligent data analysis at the network edge. However, there are also many challenges and open issues to consider, such as potential data loss, limited computing power of edge devices, hardware and ML model variations, as well as privacy and security concerns. 

Beyond these approaches, semantic and goal-oriented communication present an exciting future research direction in the field of communication systems, focusing on transmitting the meaning of data rather than raw content for improved efficiency. Semantic and goal-oriented communication is outside the scope of this paper, and we intend to explore it further in our future work.

\section*{Acknowledgements}
This work has been supported in part by Croatian Science Foundation under the project IP-2019-04-1986 \textit{IoT4us: Smart human-centric services in interoperable and decentralised IoT environments} and by the Horizon Europe WIDERA program under the grant agreement No. 101079214 \textit{(AIoTwin)}.

%
%
%
\bibliographystyle{unsrt}
\bibliography{WP2}

\begin{thebibliography}{100}

\bibitem{cisco}
Cisco.
\newblock Cisco annual internet report (2018–2023).
\newblock \url{https://www.cisco.com/c/en/us/solutions/collateral/executive-perspectives/annual-internet-report/white-paper-c11-741490.html}, 2023.
\newblock [Online; accessed 13-June-2024].

\bibitem{energyiot}
Shikhar Suryavansh, Abu Benna, Chris Guest, and Somali Chaterji.
\newblock A data-driven approach to increasing the lifetime of iot sensor nodes.
\newblock {\em Scientific Reports}, 11:22459, 11 2021.

\bibitem{iea}
IEA (International~Energy Agency).
\newblock Iea (2024), electricity 2024, iea, paris.
\newblock \url{https://www.iea.org/reports/electricity-2024}, 2024.
\newblock [Online; accessed 13-June-2024].

\bibitem{7469991}
Weisong Shi and Schahram Dustdar.
\newblock The promise of edge computing.
\newblock {\em Computer}, 49(5):78--81, 2016.

\bibitem{big_data}
Haiyan Yu, Akinola Ogbeyemi, Wenjun Lin, Jingyi He, Wei Sun, and W.J. Zhang.
\newblock A semantic model for enterprise application integration in the era of data explosion and globalisation.
\newblock {\em Enterprise Information Systems}, 17:1--23, 10 2021.

\bibitem{9063670}
Latif~U. Khan, Ibrar Yaqoob, Nguyen~H. Tran, S.~M.~Ahsan Kazmi, Tri~Nguyen Dang, and Choong~Seon Hong.
\newblock Edge-computing-enabled smart cities: A comprehensive survey.
\newblock {\em IEEE Internet of Things Journal}, 7(10):10200--10232, 2020.

\bibitem{TALAVERA2017283}
Jesús~Martín Talavera, Luis~Eduardo Tobón, Jairo~Alejandro Gómez, María~Alejandra Culman, Juan~Manuel Aranda, Diana~Teresa Parra, Luis~Alfredo Quiroz, Adolfo Hoyos, and Luis~Ernesto Garreta.
\newblock Review of iot applications in agro-industrial and environmental fields.
\newblock {\em Computers and Electronics in Agriculture}, 142:283--297, 2017.

\bibitem{inproceedings}
Zhibin Guan, Tongkai Ji, Xu~Qian, Yan Ma, and Xuehai Hong.
\newblock A survey on big data pre-processing.
\newblock In {\em 2017 5th Intl Conf on Applied Computing and Information Technology/4th Intl Conf on Computational Science/Intelligence and Applied Informatics/2nd Intl Conf on Big Data, Cloud Computing, Data Science (ACIT-CSII-BCD)}, pages 241--247, 2017.

\bibitem{article}
Laercio Pioli~Junior, Carina Dorneles, Douglas Macedo, and Mario Dantas.
\newblock An overview of data reduction solutions at the edge of iot systems: a systematic mapping of the literature.
\newblock {\em Computing}, 104:1--23, 08 2022.

\bibitem{GERODIMOS20231}
Apostolos Gerodimos, Leandros Maglaras, Mohamed~Amine Ferrag, Nick Ayres, and Ioanna Kantzavelou.
\newblock Iot: Communication protocols and security threats.
\newblock {\em Internet of Things and Cyber-Physical Systems}, 3:1--13, 2023.

\bibitem{NASSRA2023100806}
Ihab Nassra and Juan~V. Capella.
\newblock Data compression techniques in iot-enabled wireless body sensor networks: A systematic literature review and research trends for qos improvement.
\newblock {\em Internet of Things}, 23:100806, 2023.

\bibitem{aggregation_survey}
Beneyaz~Ara Begum, Satyanarayana~V Nandury, et~al.
\newblock A survey of data aggregation protocols for energy conservation in wsn and iot.
\newblock {\em Wireless Communications and Mobile Computing}, 2022, 2022.

\bibitem{rfc7228}
Carsten Bormann, Mehmet Ersue, and Ari Keränen.
\newblock {Terminology for Constrained-Node Networks}, 2014.

\bibitem{liu2024multiobjectiveoptimizationdatacollection}
Lingling Liu, Aimin Wang, Geng Sun, Jiahui Li, Hongyang Pan, and Tony Q.~S. Quek.
\newblock Multi-objective optimization for data collection in uav-assisted agricultural iot, 2024.

\bibitem{9776521}
Lingling Liu, Aimin Wang, Geng Sun, and Jiahui Li.
\newblock Multiobjective optimization for improving throughput and energy efficiency in uav-enabled iot.
\newblock {\em IEEE Internet of Things Journal}, 9(20):20763--20777, 2022.

\bibitem{9066337}
Miljan Sikimić, Momčilo Amović, Vladimir Vujović, Bojan Suknović, and Dragan Manjak.
\newblock An overview of wireless technologies for iot network.
\newblock In {\em 2020 19th International Symposium INFOTEH-JAHORINA (INFOTEH)}, pages 1--6, 2020.

\bibitem{9706334}
Arup Barua, Md~Abdullah Al~Alamin, Md.~Shohrab Hossain, and Ekram Hossain.
\newblock Security and privacy threats for bluetooth low energy in iot and wearable devices: A comprehensive survey.
\newblock {\em IEEE Open Journal of the Communications Society}, 3:251--281, 2022.

\bibitem{ZOHOURIAN2023100791}
Alireza Zohourian, Sajjad Dadkhah, Euclides Carlos~Pinto Neto, Hassan Mahdikhani, Priscilla~Kyei Danso, Heather Molyneaux, and Ali~A. Ghorbani.
\newblock Iot zigbee device security: A comprehensive review.
\newblock {\em Internet of Things}, 22:100791, 2023.

\bibitem{ergen2004zigbee}
Sinem~Coleri Ergen.
\newblock Zigbee/ieee 802.15. 4 summary.
\newblock {\em UC Berkeley, September}, 10(17):11, 2004.

\bibitem{6616827}
Artem Dementyev, Steve Hodges, Stuart Taylor, and Joshua Smith.
\newblock Power consumption analysis of bluetooth low energy, zigbee and ant sensor nodes in a cyclic sleep scenario.
\newblock In {\em 2013 IEEE International Wireless Symposium (IWS)}, pages 1--4, 2013.

\bibitem{8847082}
Anirudh Ganji, Griffin Page, and Muhammad Shahzad.
\newblock Characterizing the performance of wifi in dense iot deployments.
\newblock In {\em 2019 28th International Conference on Computer Communication and Networks (ICCCN)}, pages 1--9, 2019.

\bibitem{s18113995}
Jetmir Haxhibeqiri, Eli De~Poorter, Ingrid Moerman, and Jeroen Hoebeke.
\newblock A survey of lorawan for iot: From technology to application.
\newblock {\em Sensors}, 18(11), 2018.

\bibitem{SINHA201714}
Rashmi~Sharan Sinha, Yiqiao Wei, and Seung-Hoon Hwang.
\newblock A survey on lpwa technology: Lora and nb-iot.
\newblock {\em ICT Express}, 3(1):14--21, 2017.

\bibitem{7962670}
Almudena~Díaz Zayas and Pedro Merino.
\newblock The 3gpp nb-iot system architecture for the internet of things.
\newblock In {\em 2017 IEEE International Conference on Communications Workshops (ICC Workshops)}, pages 277--282, 2017.

\bibitem{MOHAN2023103723}
Vamshi~Sunku Mohan, Sriram Sankaran, Priyadarsi Nanda, and Krishnashree Achuthan.
\newblock Enabling secure lightweight mobile narrowband internet of things (nb-iot) applications using blockchain.
\newblock {\em Journal of Network and Computer Applications}, page 103723, 2023.

\bibitem{10.1145/3551663.3558596}
Richard Verhoeven, Stash Kempinski, and Nirvana Meratnia.
\newblock Performance evaluation of wi-fi halow, nb-iot and lora for smart city applications.
\newblock In {\em Proceedings of the 19th ACM International Symposium on Performance Evaluation of Wireless Ad Hoc, Sensor, and Ubiquitous Networks}, PE-WASUN '22, page 17–24, New York, NY, USA, 2022. Association for Computing Machinery.

\bibitem{9711564}
Mojtaba Vaezi, Amin Azari, Saeed~R. Khosravirad, Mahyar Shirvanimoghaddam, M.~Mahdi Azari, Danai Chasaki, and Petar Popovski.
\newblock Cellular, wide-area, and non-terrestrial iot: A survey on 5g advances and the road toward 6g.
\newblock {\em IEEE Communications Surveys and Tutorials}, 24(2):1117--1174, 2022.

\bibitem{BLErange}
Tong Ji, Wenhao Li, Xiangcheng Zhu, and Mengshuang Liu.
\newblock Survey on indoor fingerprint localization for ble.
\newblock In {\em 2022 IEEE 6th Information Technology and Mechatronics Engineering Conference (ITOEC)}, volume~6, pages 129--134, 2022.

\bibitem{ZigBeeRange}
Ievgeniia Kuzminykh, Arkadii Snihurov, and Anders Carlsson.
\newblock Testing of communication range in zigbee technology.
\newblock In {\em 2017 14th International Conference The Experience of Designing and Application of CAD Systems in Microelectronics (CADSM)}, pages 133--136, 2017.

\bibitem{WiFiRange}
Shailandra Kaushik.
\newblock An overview of technical aspect for wifi networks technology.
\newblock {\em International Journal of Electronics and Computer Science Engineering}, 1, 04 2012.

\bibitem{LoRaWANRange}
Ferran Adelantado, Xavier Vilajosana, Pere Tuset-Peiro, Borja Martinez, Joan Melia-Segui, and Thomas Watteyne.
\newblock Understanding the limits of lorawan.
\newblock {\em IEEE Communications Magazine}, 55(9):34--40, 2017.

\bibitem{4460126}
Jin-Shyan Lee, Yu-Wei Su, and Chung-Chou Shen.
\newblock A comparative study of wireless protocols: Bluetooth, uwb, zigbee, and wi-fi.
\newblock In {\em IECON 2007 - 33rd Annual Conference of the IEEE Industrial Electronics Society}, pages 46--51, 2007.

\bibitem{8053135}
Abdullah Kurtoglu, Joan Carletta, and Kye-Shin Lee.
\newblock Energy consumption in long-range linear wireless sensor networks using lorawan and zigbee.
\newblock In {\em 2017 IEEE 60th International Midwest Symposium on Circuits and Systems (MWSCAS)}, pages 1163--1167, 2017.

\bibitem{7334098}
Jin-Shyan Lee, Ming-Feng Dong, and Yuan-Heng Sun.
\newblock A preliminary study of low power wireless technologies: Zigbee and bluetooth low energy.
\newblock In {\em 2015 IEEE 10th Conference on Industrial Electronics and Applications (ICIEA)}, pages 135--139, 2015.

\bibitem{articlewbfb}
Jiun-ren Lin, Timothy Talty, and O.K. Tonguz.
\newblock On the potential of bluetooth low energy technology in vehicular applications.
\newblock {\em IEEE Communications Magazine}, 53:267--275, 02 2015.

\bibitem{9068491}
Massimo Ballerini, Tommaso Polonelli, Davide Brunelli, Michele Magno, and Luca Benini.
\newblock Nb-iot versus lorawan: An experimental evaluation for industrial applications.
\newblock {\em IEEE Transactions on Industrial Informatics}, 16(12):7802--7811, 2020.

\bibitem{8614763}
F.~John Dian, Amirhossein Yousefi, and Sungjoon Lim.
\newblock A practical study on bluetooth low energy (ble) throughput.
\newblock In {\em 2018 IEEE 9th Annual Information Technology, Electronics and Mobile Communication Conference (IEMCON)}, pages 768--771, 2018.

\bibitem{5942102}
C.~Muthu Ramya, M~Shanmugaraj, and R~Prabakaran.
\newblock Study on zigbee technology.
\newblock In {\em 2011 3rd International Conference on Electronics Computer Technology}, volume~6, pages 297--301, 2011.

\bibitem{fi11010016}
Luiz Oliveira, Joel J. P.~C. Rodrigues, Sergei~A. Kozlov, Ricardo A.~L. Rabêlo, and Victor Hugo C.~de Albuquerque.
\newblock Mac layer protocols for internet of things: A survey.
\newblock {\em Future Internet}, 11(1), 2019.

\bibitem{DONTA2022727}
Praveen~Kumar Donta, Satish~Narayana Srirama, Tarachand Amgoth, and Chandra Sekhara~Rao Annavarapu.
\newblock Survey on recent advances in iot application layer protocols and machine learning scope for research directions.
\newblock {\em Digital Communications and Networks}, 8(5):727--744, 2022.

\bibitem{8088251}
Nitin Naik.
\newblock Choice of effective messaging protocols for iot systems: Mqtt, coap, amqp and http.
\newblock In {\em 2017 IEEE International Systems Engineering Symposium (ISSE)}, pages 1--7, 2017.

\bibitem{7745303}
Muneer~Bani Yassein, Mohammed~Q. Shatnawi, and Dua' Al-zoubi.
\newblock Application layer protocols for the internet of things: A survey.
\newblock In {\em 2016 International Conference on Engineering and MIS (ICEMIS)}, pages 1--4, 2016.

\bibitem{9247996}
Biswajeeban Mishra and Attila Kertesz.
\newblock The use of mqtt in m2m and iot systems: A survey.
\newblock {\em IEEE Access}, 8:201071--201086, 2020.

\bibitem{IQBAL2023109640}
Faheem Iqbal, Moneeb Gohar, Hanen Karamti, Walid Karamti, Seok-Joo Koh, and Jin-Ghoo Choi.
\newblock Use of quic for amqp in iot networks.
\newblock {\em Computer Networks}, 225:109640, 2023.

\bibitem{9023812}
Nguyen~Quoc Uy and Vu~Hoai Nam.
\newblock A comparison of amqp and mqtt protocols for internet of things.
\newblock In {\em 2019 6th NAFOSTED Conference on Information and Computer Science (NICS)}, pages 292--297, 2019.

\bibitem{8914552}
Thays Moraes, Bruno Nogueira, Victor Lira, and Eduardo Tavares.
\newblock Performance comparison of iot communication protocols.
\newblock In {\em 2019 IEEE International Conference on Systems, Man and Cybernetics (SMC)}, pages 3249--3254, 2019.

\bibitem{9559032}
Cavide~Balkı Gemirter, Cagatay Senturca, and Sebnem Baydere.
\newblock A comparative evaluation of amqp, mqtt and http protocols using real-time public smart city data.
\newblock In {\em 2021 6th International Conference on Computer Science and Engineering (UBMK)}, pages 542--547, 2021.

\bibitem{fARHAN}
Laith Farhan, Rupak Kharel, Omprakash Kaiwartya, Marcela Quiroz-Castellanos, Ali Alissa, and Mohamed Abdulsalam.
\newblock A concise review on internet of things (iot) -problems, challenges and opportunities.
\newblock In {\em 2018 11th International Symposium on Communication Systems, Networks and Digital Signal Processing (CSNDSP)}, pages 1--6, 2018.

\bibitem{Lossy_review_iot}
Juan David~Arias Correa, Alex Sandro~Roschildt Pinto, and Carlos Montez.
\newblock Lossy data compression for iot sensors: A review.
\newblock {\em Internet of Things}, 19:100516, 2022.

\bibitem{gubitci}
Mohammad Hosseini.
\newblock A survey of data compression algorithms and their applications.
\newblock {\em Network Systems Laboratory, School of Computing Science, Simon Fraser University, BC, Canada}, 2012.

\bibitem{kompresija_2}
Santosh Pattar, Rajkumar Buyya, K.~R. Venugopal, S.~S. Iyengar, and L.~M. Patnaik.
\newblock Searching for the iot resources: Fundamentals, requirements, comprehensive review, and future directions.
\newblock {\em IEEE Communications Surveys and Tutorials}, 20(3):2101--2132, 2018.

\bibitem{lossless}
Emanuel Guberović, Fran Krišto, Petar Krivić, and Igor Čavrak.
\newblock Assessing compression algorithms on iot sensor nodes.
\newblock In {\em 2019 42nd International Convention on Information and Communication Technology, Electronics and Microelectronics (MIPRO)}, pages 913--918, 2019.

\bibitem{two_tier}
Ali Kadhum~M. Al-Qurabat, Chady Abou~Jaoude, and Ali~Kadhum Idrees.
\newblock Two tier data reduction technique for reducing data transmission in iot sensors.
\newblock In {\em 2019 15th International Wireless Communications and Mobile Computing Conference (IWCMC)}, pages 168--173, 2019.

\bibitem{sprintz}
Davis~W. Blalock, Samuel Madden, and John~V. Guttag.
\newblock Sprintz: Time series compression for the internet of things.
\newblock {\em Proc. ACM Interact. Mob. Wearable Ubiquitous Technol.}, 2:93:1--93:23, 2018.

\bibitem{delta}
Biljana~Risteska Stojkoska and Zoran Nikolovski.
\newblock Data compression for energy efficient iot solutions.
\newblock In {\em 2017 25th Telecommunication Forum (TELFOR)}, pages 1--4, 2017.

\bibitem{lossy}
Uthayakumar Jayasankar, Vengattaraman Thirumal, and Dhavachelvan Ponnurangam.
\newblock A survey on data compression techniques: From the perspective of data quality, coding schemes, data type and applications.
\newblock {\em Journal of King Saud University - Computer and Information Sciences}, 33(2):119--140, 2021.

\bibitem{agg_sur}
Behrouz Pourghebleh and Nima~Jafari Navimipour.
\newblock Data aggregation mechanisms in the internet of things: A systematic review of the literature and recommendations for future research.
\newblock {\em Journal of Network and Computer Applications}, 97:23--34, 2017.

\bibitem{agg_comp}
M~Amarlingam, Pradeep~Kumar Mishra, Pachamuthu Rajalakshmi, Sumohana~S Channappayya, and Challa~S Sastry.
\newblock Novel light weight compressed data aggregation using sparse measurements for iot networks.
\newblock {\em Journal of Network and Computer Applications}, 121:119--134, 2018.

\bibitem{agg_mech}
Ali~Kadhum Idrees, Ali Kadhum~M Al-Qurabat, Chady Abou~Jaoude, and Wathiq~Laftah Al-Yaseen.
\newblock Integrated divide and conquer with enhanced k-means technique for energy-saving data aggregation in wireless sensor networks.
\newblock In {\em 2019 15th International wireless communications and mobile computing conference (IWCMC)}, pages 973--978. IEEE, 2019.

\bibitem{added}
Junwei Wang, Dong Liu, W.H. Ip, Wenjun Zhang, and Ralph Deters.
\newblock Integration of system-dynamics, aspect-programming, and object-orientation in system information modeling.
\newblock {\em Industrial Informatics, IEEE Transactions on}, 10:847--853, 05 2014.

\bibitem{2020edge}
Huihui Xue, Bi~Huang, Mingming Qin, Hua Zhou, and Hongji Yang.
\newblock Edge computing for internet of things: A survey.
\newblock In {\em 2020 International Conferences on Internet of Things (iThings) and IEEE Green Computing and Communications (GreenCom) and IEEE Cyber, Physical and Social Computing (CPSCom) and IEEE Smart Data (SmartData) and IEEE Congress on Cybermatics (Cybermatics)}, pages 755--760. IEEE, 2020.

\bibitem{OLIVEIRA2024101153}
n~Oliveira Frankli, Costa Daniel~G., Assis Flávio, and Silva Ivanovitch.
\newblock Internet of intelligent things: A convergence of embedded systems, edge computing and machine learning.
\newblock {\em Internet of Things}, 26:101153, 2024.

\bibitem{dias2016survey}
Gabriel~Martins Dias, Boris Bellalta, and Simon Oechsner.
\newblock A survey about prediction-based data reduction in wireless sensor networks.
\newblock {\em ACM Computing Surveys (CSUR)}, 49(3):1--35, 2016.

\bibitem{hussein2022distributed}
Ahmed~Mohammed Hussein, Ali~Kadhum Idrees, and Raphael Couturier.
\newblock Distributed energy-efficient data reduction approach based on prediction and compression to reduce data transmission in iot networks.
\newblock {\em International Journal of Communication Systems}, 35(15):e5282, 2022.

\bibitem{10143415}
Made Adi~Paramartha Putra, Ade~Pitra Hermawan, Dong-Seong Kim, and Jae-Min Lee.
\newblock Data prediction-based energy-efficient architecture for industrial iot.
\newblock {\em IEEE Sensors Journal}, 23(14):15856--15866, 2023.

\bibitem{dual_prediction}
Haibin Wang, Zaid Yemeni, Waleed Ismael, Ammar Hawbani, and Saeed Alsamhi.
\newblock A reliable and energy efficient dual prediction data reduction approach for wsns based on kalman filter.
\newblock {\em IET Communications}, page~1, 11 2021.

\bibitem{PLACZEK2024225}
Bartłomiej Płaczek.
\newblock Prediction-based data reduction with dynamic target node selection in iot sensor networks.
\newblock {\em Future Generation Computer Systems}, 152:225--238, 2024.

\bibitem{review_forecast}
Luis-Roberto J{\'a}come-Galarza, Miguel-Andr{\'e}s Realpe-Robalino, Jonathan Paillacho-Corredores, and Jos{\'e}-Leonardo Benavides-Maldonado.
\newblock Time series in sensor data using state-of-the-art deep learning approaches: A systematic literature review.
\newblock In {\'A}lvaro Rocha, Paulo~Carlos L{\'o}pez-L{\'o}pez, and Juan~Pablo Salgado-Guerrero, editors, {\em Communication, Smart Technologies and Innovation for Society}, pages 503--514, Singapore, 2022. Springer Singapore.

\bibitem{jarwan2019data}
Abdallah Jarwan, Ayman Sabbah, and Mohamed Ibnkahla.
\newblock Data transmission reduction schemes in wsns for efficient iot systems.
\newblock {\em IEEE Journal on Selected Areas in Communications}, 37(6):1307--1324, 2019.

\bibitem{model_compression}
Zhuo Li, Hengyi Li, and Lin Meng.
\newblock Model compression for deep neural networks: A survey.
\newblock {\em Computers}, 12(3), 2023.

\bibitem{cheng2023surveydeepneuralnetwork}
Hongrong Cheng, Miao Zhang, and Javen~Qinfeng Shi.
\newblock A survey on deep neural network pruning-taxonomy, comparison, analysis, and recommendations, 2023.

\bibitem{Hernandez2024}
Nicolás Hernández, Francisco Almeida, and Vicente Blanco.
\newblock Optimizing convolutional neural networks for iot devices: performance and energy efficiency of quantization techniques.
\newblock {\em The Journal of Supercomputing}, 80(9):12686--12705, 2024.

\bibitem{knowledge}
Xiaohan Xu, Ming Li, Chongyang Tao, Tao Shen, Reynold Cheng, Jinyang Li, Can Xu, Dacheng Tao, and Tianyi Zhou.
\newblock A survey on knowledge distillation of large language models, 2024.

\bibitem{SADRI2022100629}
Ali~Akbar Sadri, Amir~Masoud Rahmani, Morteza Saberikamarposhti, and Mehdi Hosseinzadeh.
\newblock Data reduction in fog computing and internet of things: A systematic literature survey.
\newblock {\em Internet of Things}, 20:100629, 2022.

\bibitem{compression_iot}
Tulika Bose, Soma Bandyopadhyay, Sudhir Kumar, Abhijan Bhattacharyya, and Arpan Pal.
\newblock Signal characteristics on sensor data compression in iot - an investigation.
\newblock In {\em 2016 IEEE International Conference on Sensing, Communication and Networking (SECON Workshops)}, pages 1--6. IEEE, 6 2016.

\bibitem{gateway}
Cinthya~M. França, Rodrigo~S. Couto, and Pedro~B. Velloso.
\newblock Missing data imputation in internet of things gateways.
\newblock {\em Information}, 12(10), 2021.

\bibitem{missing}
Cinthya~M. França, Rodrigo~S. Couto, and Pedro~B. Velloso.
\newblock Data imputation on iot gateways using machine learning.
\newblock In {\em 2021 19th Mediterranean Communication and Computer Networking Conference (MedComNet)}, pages 1--8, 2021.

\bibitem{fi16010029}
Valeriy Ivanov and Maxim Tereshonok.
\newblock Cross-layer methods for ad hoc networks—review and classification.
\newblock {\em Future Internet}, 16(1), 2024.

\bibitem{cl}
Sultan Chowdhury and Ashraf Hossain.
\newblock Different energy saving schemes in wireless sensor networks: A survey.
\newblock {\em Wireless Personal Communications}, 114, 10 2020.

\bibitem{articleResAllocation}
Leonidas Georgiadis, Michael Neely, and Leandros Tassiulas.
\newblock Resource allocation and cross-layer control in wireless networks.
\newblock {\em Foundations and Trends in Networking}, 1, 01 2006.

\bibitem{7986381}
Asma Messaoudi, Rabiaa Elkamel, Abdelhamid Helali, and Ridha Bouallegue.
\newblock Cross-layer based routing protocol for wireless sensor networks using a fuzzy logic module.
\newblock In {\em 2017 13th International Wireless Communications and Mobile Computing Conference (IWCMC)}, pages 764--769, 2017.

\bibitem{4205060}
C.~Wang, B.~Li, K.~Sohraby, M.~Daneshmand, and Y.~Hu.
\newblock Upstream congestion control in wireless sensor networks through cross-layer optimization.
\newblock {\em IEEE Journal on Selected Areas in Communications}, 25(4):786--795, 2007.

\bibitem{10.1007/978-981-19-2719-5_62}
K.~Swaminathan, Vijay Ravindran, Ramprakash Ponraj, and R.~Satheesh.
\newblock A smart energy optimization and collision avoidance routing strategy for iot systems in the wsn domain.
\newblock In Brijesh Iyer, Tom Crick, and Sheng-Lung Peng, editors, {\em Applied Computational Technologies}, pages 655--663, Singapore, 2022. Springer Nature Singapore.

\bibitem{HASAN2022101099}
Nadine Hasan, Ayaskanta Mishra, and Arun~Kumar Ray.
\newblock Fuzzy logic based cross-layer design to improve quality of service in mobile ad-hoc networks for next-gen cyber physical system.
\newblock {\em Engineering Science and Technology, an International Journal}, 35:101099, 2022.

\bibitem{parween2020review}
Sultana Parween and Syed~Zeeshan Hussain.
\newblock A review on cross-layer design approach in wsn by different techniques.
\newblock {\em Adv. Sci. Technol. Eng. Syst}, 5(4):741--754, 2020.

\bibitem{s23031639}
Amira Bourechak, Ouarda Zedadra, Mohamed~Nadjib Kouahla, Antonio Guerrieri, Hamid Seridi, and Giancarlo Fortino.
\newblock At the confluence of artificial intelligence and edge computing in iot-based applications: A review and new perspectives.
\newblock {\em Sensors}, 23(3), 2023.

\bibitem{SINGH202371}
Raghubir Singh and Sukhpal~Singh Gill.
\newblock Edge ai: A survey.
\newblock {\em Internet of Things and Cyber-Physical Systems}, 3:71--92, 2023.

\bibitem{vehicles}
Anushka Biswas and Hwang-Cheng Wang.
\newblock Autonomous vehicles enabled by the integration of iot, edge intelligence, 5g, and blockchain.
\newblock {\em Sensors}, 23(4), 2023.

\bibitem{DBLP:conf/icra/IchnowskiCDADVJZLXBKSGG23}
Jeffrey Ichnowski, Kaiyuan Chen, Karthik Dharmarajan, Simeon Adebola, Michael Danielczuk, Victor~Mayoral Vilches, Nikhil Jha, Hugo Zhan, Edith LLontop, Derek Xu, Camilo Buscaron, John Kubiatowicz, Ion Stoica, Joseph Gonzalez, and Ken Goldberg.
\newblock Fogros2: An adaptive platform for cloud and fog robotics using {ROS} 2.
\newblock In {\em {IEEE} International Conference on Robotics and Automation, {ICRA} 2023, London, UK, May 29 - June 2, 2023}, pages 5493--5500. {IEEE}, 2023.

\bibitem{DBLP:journals/iotj/HuangZCLZY22}
Peng Huang, Liekang Zeng, Xu~Chen, Ke~Luo, Zhi Zhou, and Shuai Yu.
\newblock Edge robotics: Edge-computing-accelerated multirobot simultaneous localization and mapping.
\newblock {\em {IEEE} Internet Things J.}, 9(15):14087--14102, 2022.

\bibitem{Jingao:2022}
Jingao Xu, Hao Cao, Zheng Yang, Longfei Shangguan, Jialin Zhang, Xiaowu He, and Yunhao Liu.
\newblock {SwarmMap}: Scaling up real-time collaborative visual {SLAM} at the edge.
\newblock In {\em 19th USENIX Symposium on Networked Systems Design and Implementation (NSDI 22)}, pages 977--993, Renton, WA, April 2022. USENIX Association.

\bibitem{lajoieSwarmSLAM}
Pierre-Yves Lajoie and Giovanni Beltrame.
\newblock Swarm-slam: Sparse decentralized collaborative simultaneous localization and mapping framework for multi-robot systems.
\newblock {\em IEEE Robotics and Automation Letters}, 9(1):475--482, 2024.

\bibitem{SOORI2023192}
Mohsen Soori, Behrooz Arezoo, and Roza Dastres.
\newblock Internet of things for smart factories in industry 4.0, a review.
\newblock {\em Internet of Things and Cyber-Physical Systems}, 3:192--204, 2023.

\bibitem{electronics12041027}
Abdulrahman~K. Alnaim and Ahmed~M. Alwakeel.
\newblock Machine-learning-based iot–edge computing healthcare solutions.
\newblock {\em Electronics}, 12(4), 2023.

\bibitem{cilicPerf}
Ivan Čilić, Petar Krivić, Ivana Podnar~Žarko, and Mario Kušek.
\newblock Performance evaluation of container orchestration tools in edge computing environments.
\newblock {\em Sensors}, 23(8), 2023.

\bibitem{ZHANG2021106775}
Chen Zhang, Yu~Xie, Hang Bai, Bin Yu, Weihong Li, and Yuan Gao.
\newblock A survey on federated learning.
\newblock {\em Knowledge-Based Systems}, 216:106775, 2021.

\bibitem{8951147}
Vittorio Mazzia, Aleem Khaliq, Francesco Salvetti, and Marcello Chiaberge.
\newblock Real-time apple detection system using embedded systems with hardware accelerators: An edge ai application.
\newblock {\em IEEE Access}, 8:9102--9114, 2020.

\bibitem{liu2022bringing}
Di~Liu, Hao Kong, Xiangzhong Luo, Weichen Liu, and Ravi Subramaniam.
\newblock Bringing ai to edge: From deep learning’s perspective.
\newblock {\em Neurocomputing}, 485:297--320, 2022.

\bibitem{li2023model}
Zhuo Li, Hengyi Li, and Lin Meng.
\newblock Model compression for deep neural networks: A survey.
\newblock {\em Computers}, 12(3):60, 2023.

\bibitem{liu_hiear}
Lumin Liu, Jun Zhang, S.H. Song, and Khaled~B. Letaief.
\newblock Client-edge-cloud hierarchical federated learning.
\newblock In {\em Proc. IEEE ICC}, 2020.

\bibitem{s23031279}
Chellammal Surianarayanan, John~Jeyasekaran Lawrence, Pethuru~Raj Chelliah, Edmond Prakash, and Chaminda Hewage.
\newblock A survey on optimization techniques for edge artificial intelligence (ai).
\newblock {\em Sensors}, 23(3), 2023.

\bibitem{flhealth}
Anichur Rahman, Md.~Sazzad Hossain, Ghulam Muhammad, Dipanjali Kundu, Tanoy Debnath, Muaz Rahman, Md~Khan, Prayag Tiwari, and Shahab S.~Band.
\newblock Federated learning-based ai approaches in smart healthcare: concepts, taxonomies, challenges and open issues, 08 2022.

\bibitem{DBLP:journals/corr/abs-2305-19831}
Kok{-}Seng Wong, Manh Nguyen{-}Duc, Khiem Le{-}Huy, Long Ho{-}Tuan, Cuong Do{-}Danh, and Danh~Le Phuoc.
\newblock An empirical study of federated learning on iot-edge devices: Resource allocation and heterogeneity.
\newblock {\em CoRR}, abs/2305.19831, 2023.

\bibitem{DBLP:journals/corr/abs-2405-20431}
Khiem Le, Nhan Luong{-}Ha, Manh Nguyen{-}Duc, Danh~Le Phuoc, Cuong Do, and Kok{-}Seng Wong.
\newblock Exploring the practicality of federated learning: {A} survey towards the communication perspective.
\newblock {\em CoRR}, abs/2405.20431, 2024.

\bibitem{levisse}
Alexandre Levisse, Marco Rios, W.-A. Simon, P.-E. Gaillardon, and D.~Atienza.
\newblock Functionality enhanced memories for edge-ai embedded systems.
\newblock In {\em 2019 19th Non-Volatile Memory Technology Symposium (NVMTS)}, pages 1--4, 2019.

\bibitem{9094236}
Deyin Liu, Xu~Chen, Zhi Zhou, and Qing Ling.
\newblock Hiertrain: Fast hierarchical edge ai learning with hybrid parallelism in mobile-edge-cloud computing.
\newblock {\em IEEE Open Journal of the Communications Society}, 1:634--645, 2020.

\end{thebibliography}
%
%
%
%
%
\end{document}